\documentclass[10pt,a4paper]{article}
\usepackage{amsmath, amssymb, amsthm}
\usepackage{fullpage}
\usepackage{sgame}
\usepackage{color}
\usepackage{tikz}
\usetikzlibrary{positioning,shadows,shapes,backgrounds,fit}
\usepackage{authblk}
\usepackage[numbers]{natbib}

\usepackage{ifpdf}
\ifpdf
  \usepackage[pdftex]{hyperref}
\else
  \usepackage[hypertex]{hyperref}
\fi

\newtheorem{definition}{Definition}[section]
\newtheorem{theorem}[definition]{Theorem}
\newtheorem{lemma}[definition]{Lemma}

\newtheorem{obs}[definition]{Observation}
\newtheorem{cor}[definition]{Corollary}
\newtheorem{fact}{Fact}

\newcommand{\nlemma}[3]{\newtheorem*{lemma#1}{Lemma~#2}\begin{lemma#1}#3\end{lemma#1}}

\newcommand{\Prob}[2]{\mathbf{P}_{#1} \left( #2 \right)}
\newcommand{\Expec}[2]{\mathbf{E}_{#1} \left[ #2 \right]}

\newcommand{\x}{{\mathbf{x}}}

\newcommand{\y}{{\mathbf y}}

\newcommand{\p}{{\mathbf p}}
\newcommand{\m}{{\mathbf m}}
\newcommand{\+}{\boldsymbol{+}}
\newcommand{\meno}{\boldsymbol{-}}

\newcommand{\tm}{{t_\text{\rm mix}}}

\renewcommand{\leq}{\leqslant}
\renewcommand{\geq}{\geqslant}

\renewcommand{\epsilon}{\varepsilon}
\renewcommand{\mod}{\text{ {\rm mod} }}

\newcommand{\tv}[1]{\left\|#1\right\|_{\rm TV}}
\newcommand{\pmt}[3]{t_{#1}^{#2}(#3)}

\newcommand{\Bn}{B}               %nel libro di Peres e' \Phi
\newcommand{\Par}[1]{p(#1)}

\newcommand{\OO}{{\mathcal O}}

\newcommand{\negl}[1]{{\sf negl}\left(#1\right)}
\newcommand{\poly}{{\sf poly}}

\newcommand{\ignore}[1]{}

\begin{document}

\title{\textbf{Metastability of Logit Dynamics for Coordination Games}\thanks{A preliminary version of this paper appeared in the Proceedings of the Twenty-Third Annual ACM-SIAM Symposium on Discrete Algorithms (SODA'12) \cite{AulettaFPP12}. Part of this work was done while the third author was at Universit\`a di Salerno}}
\author[1]{Vincenzo Auletta}
\author[1]{Diodato Ferraioli}
\author[2]{Francesco Pasquale}
\author[1]{Giuseppe Persiano}
\affil[1]{Universit\`a di Salerno}
\affil[2]{Universit\`a di Roma ``Tor Vergata''}
\date{}

\maketitle

\begin{abstract}
Logit Dynamics [Blume, Games and Economic Behavior, 1993] are randomized best response dynamics for strategic games: at every time step a player is selected uniformly at random and she chooses a new strategy according to a probability distribution biased toward strategies promising higher payoffs. This process defines an ergodic Markov chain, over the set of strategy profiles of the game, whose unique stationary distribution is the long-term equilibrium concept for the game. However, when the mixing time of the chain is large (e.g., exponential in the number of players), the stationary distribution loses its appeal as equilibrium concept, and the transient phase of the Markov chain becomes important. It can happen that the chain is ``metastable'', i.e., on a time-scale shorter than the mixing time, it stays close to some probability distribution over the state space, while in a time-scale multiple of the mixing time it jumps from one distribution to another.

In this paper we give a quantitative definition of ``metastable probability distributions'' for a Markov chain and we study the metastability of the logit dynamics for some classes of coordination games.
We first consider a pure $n$-player coordination game that highlights the distinctive features of our metastability notion based on distributions.
Then, we study coordination games on the clique without a risk-dominant strategy (which are equivalent to the well-known Glauber dynamics for the Curie-Weiss model)
and coordination games on a ring (both with and without risk-dominant strategy).

\end{abstract}

\newpage

\section{Introduction}\label{sec::intro}
Complex systems consist of a large number of components that interact according to simple rules at small scale and, despite of
this, exhibit complex large scale behaviors.
Complex systems are ubiquitous and some examples can be found in Economics (e.g., the market),
Physics (e.g., ideal gases, spin systems), Biology (e.g., evolution of life) and
Computer Science (e.g., Internet and social networks).
Analyzing, understanding how such systems evolve, and
predicting their {long-term behaviour} %%PINO: future states
is a major research endeavor.

In this paper we focus on {\em selfish} systems in which the components
(called the \emph{players}) are selfish agents, each one with a set of possible actions or
\emph{strategies} trying to maximize her own payoff.
The payoff obtained by each player depends not only on her decision but also
on the decisions of the other players.
We study specific dynamics, the {\em logit} dynamics
{(first studied by Blume \citep{blumeGEB93}) } %%PINO
and consider as solution concept their equilibrium states.
Logit dynamics are a type of {\em noisy} best response dynamics that
model in a clean and tractable way the limited knowledge
(or bounded rationality) of the players in terms of a parameter $\beta$
(in similar models studied in Physics,
$\beta$ is the inverse of the temperature).
Intuitively, a low value of $\beta$ (that is, high temperature and entropy)
represents the situation where players
choose their strategies ``nearly at random'';
a high value of $\beta$ (that is, low temperature and entropy)
represents the situation where players
pick the strategies yielding high payoff with
higher probability.
It is well known that, for every strategic game, these dynamics induce a Markov chain
with a {\em unique} stationary distribution (the Markov chain is ergodic).
Thus no equilibrium selection problem arises.
The drawback of using the stationary distribution to describe the system behavior is that
the system may take too long to converge to that distribution, unless the chain is rapidly mixing.
Logit dynamics for strategic games can be rapidly mixing or {\em not}, depending on the features
of the underlying game, the temperature/noise, and the number of players~\citep{afppSAGT10,afpppSPAA11}.
For this reason, in this work
we focus on the \emph{transient phase} of the logit dynamics and, in particular,
we try to answer the following questions
for games where logit dynamics have exponential mixing time:
is the transient phase completely \emph{chaotic} or can we spot some \emph{regularities} even at time-scales shorter than mixing time?
Ellison \cite{Ellison2000} showed that as long as the probability of suboptimal responses tends to $0$ (i.e., $\beta \rightarrow \infty$),
the logit dynamics (and many other similar noisy dynamics) usually evolve via a series of gradual steps between nearly stable states.
However, how can we describe these stable states? And what happens for lower values of $\beta$?
Obviously, at a fine-grained level any Markov chain is perfectly described by the collections
of probability distributions consisting of one distribution
for each time step and each starting profile. However, this should be contrasted with the rapidly mixing case
(i.e., a Markov chain with polynomial mixing time)
in which one can approximately describe the state of the system
(after the mixing time) using {\em one} distribution,
i.e. the stationary distribution.

Our results show that there are
games for which regularities
can be observed even in the transient phase of the logit dynamics.
In particular, we will show that it is often possible to identify a few probability distributions (the \emph{metastable} distributions) such that, depending on the starting profile, the dynamics quickly  reach one of those distributions and remain close to that one for a long time.
We can describe our results also in terms of the quantity of information
needed to predict the status of a system that evolves in time according to
the logit dynamics. We know that the long-term behavior of the system
can be compactly described in terms of a unique distribution but we have to wait a transient phase of length equal to the mixing time.
Thus, if the system is rapidly mixing
this description is significant after a short transient phase.
However, when the mixing time is super-polynomial this description becomes
significant only after a long time.
Our results show that for a large class of $n$-player games whose logit
dynamics are not rapidly mixing, the behaviour of the dynamics
(the strategies played by the $n$ players) can still be described with good approximation
and for a super-polynomial number of steps by means of
a small number of probability distributions.
This comes at the price of sacrificing a short polynomial initial transient
phase (so far we are on a par with the rapidly mixing case)
and requires a few bits of information about the starting profile
(this is not needed in the rapidly mixing case).

\paragraph{Our contribution.}
Our main contribution is the introduction
of the notion of an {\em $(\epsilon,T)$-metastable distribution} of a Markov chain
%(see Section~\ref{sec::meta})
and of the concept of {\em pseudo-mixing time} of a metastable distribution.
Roughly speaking, a distribution $\mu$ is $(\epsilon,T)$-metastable for a Markov chain
if, starting from $\mu$, the Markov chain stays at distance at most
$\epsilon$ from $\mu$ for at least $T$ steps.
The pseudo-mixing time of $\mu$ starting from a state $x$ is the number
of steps needed by the Markov chain to get $\epsilon$-close to $\mu$ when started from $x$.

In a rapidly-mixing Markov chain, after a ``short time" and regardless
of the starting state, the chain converges rapidly to the stationary
distribution and remains there.
For the case of \emph{non-}rapidly-mixing Markov chains,
we replace the notions of ``mixing time'' and ``stationary distribution''
with those of ``pseudo-mixing time'' and ``metastable distribution''.
Intuitively speaking, we would like to say that, even when the mixing time
is (prohibitively) high, there are ``few'' distributions which give us
an accurate description of the chain over a ``reasonable amount of time''.
That is, the state space $\Omega$ can be partitioned into
a small number of subsets $\Omega_1,\Omega_2,\ldots$ of ``equivalent''
states so that if the chain starts in \emph{any} of the states
in $\Omega_i$, then it will
rapidly converge to a ``metastable''  distribution $\mu_i$,
where metastable denotes the fact that the chain remains there for
``sufficiently'' long.

The similarities with stationary distributions
and mixing time are confirmed by some
of the properties that are
enjoyed by the metastable distribution and the pseudo-mixing time concepts.
In Section~\ref{sec::meta} we highlight these properties
and also the relationship of these concepts with other well-known
Markov chain measures,
such as \emph{bottleneck ratio} and \emph{hitting time}.

In order to familiarize with the concepts of metastable distribution
and pseudo-mixing time, we first apply them to a simple three-state Markov chain and
the Markov chain of the logit dynamics for $2$-player coordination games.
Then, we show the usefulness of these concepts in more involved examples.
Specifically, in Section~\ref{sec::ormeta} we consider a pure $n$-player coordination game,
in which all players would like to take the same action
and each action is equivalently valued by any player.
People forming teams, objects being categorized (e.g., movies into genres in a video store)
and firms choosing their trading locations are some examples of real-world scenarios that can be modeled by this class of games \cite{behavior}.
The behavior of the logit dynamics for this game turns out to be very interesting.
Indeed, it highlights a distinctive feature of our metastability notion based on distributions,
namely that a metastable distribution exists even if every small subset of states in the support
of that distribution is quickly left with high probability.

Finally, in Section~\ref{sec::graphmeta}, we obtain results about the metastability of the
logit dynamics for different classes of graphical coordination games.
These games are often used to model the spread of
a new technology in a social network \citep{youngPUP98,youngTR00, msFOCS09} with
the strategy of maximum potential corresponding to adopting the new technology;
players prefer to choose the same technology as their neighbors
and the new technology is at least as preferable as the old one.
Research along this direction has mainly focused on analyzing
how the features of the social network affect the spread of innovations.
This research line has been initiated by \citet{ellisonECO93},
that considers two extremal network topologies, the clique and the
ring\footnote{We point out that the dynamics studied by \citet{ellisonECO93} are
slightly different from the logit dynamics, that have been instead adopted by
the later works \citep{youngPUP98,youngTR00, msFOCS09}.}.

In this work we follow \citet{ellisonECO93} and study
the case in which social interaction between the players is
described by the clique and ring topologies.
Specifically, in Section~\ref{sec::isingmeta},
we analyze the metastable distributions
for a special graphical coordination game
embedding the
Ising model on the complete graph,
also known as the Curie-Weiss model.
It has been studied in the context of population protocols \cite{sandholmMIT10} and used by physicists to model the interaction between magnets in a ferro-magnetic system~\citep{Mart1999}.
Indeed, it is known~\citep{Galam2010481,msFOCS09,afpppSPAA11} that this model can be seen as a game played by magnets and, in particular, the Glauber dynamics for the Gibbs measure on the Ising model are equivalent to the logit dynamics for this game where $\beta$ is exactly the inverse of the temperature. The mixing time of these dynamics is known to be exponential for every $\beta > 1/n$.
For this model, we show that distributions where all magnets have the same magnetization are $(1/n, t)$-metastable for $t$ greater than any polynomial when $\beta=\omega(\log n / n)$. Moreover we show that the pseudo-mixing time of these distributions is polynomial when the dynamics start from a profile where the difference in the number of positive and negative magnets is large.

In Section~\ref{sec::coordmeta} we focus on the ring topology and
we show that for every starting profile there is a metastable distribution
and the dynamics approach it in a polynomial number of steps.
The metastable distributions that arise in this case consist of a mixture of the
extreme distributions that assign probability $1$ to profiles in which all nodes
adopt the same strategy.
In the case of a symmetric coordination game (the two strategies are equally liked)
the mixture that arises has coefficients that are almost equal to the fraction
of nodes adopting each of the two strategies.
If the game is asymmetric, it turns out that
coefficients are biased towards the profile
in which all nodes adopt the most preferred strategy.
In particular, this coefficient gets very close to $1$ even if in the initial profile
there is a constant number (depending on the metastability parameter $\epsilon$)
of nodes with this strategy.

\paragraph{Other equilibrium notions.}
We find illustrative to compare the concepts studied in this paper
(the logit dynamics and their metastable distributions) with the notion of Nash equilibrium.
Similar comparison can be made with other solution concepts from Game Theory
(correlated equilibria, sink equilibria).
The notion of a Nash equilibrium has been extensively studied
as a solution concept for predicting the behavior of selfish players.
Indeed, if the players happen to be in a (pure) Nash equilibrium any
sequence of selfish best response (i.e., utility improving) moves keeps the players in the same state.
Unfortunately, the theory of Nash equilibria does not explain
how a Nash equilibrium is reached if players do not start from one and,
in case multiple equilibria exist, does not say which equilibrium
is selected (about this important issue see \citep{hsMIT88}).
Even tough it is not hard to see that sequences of best response moves may
reach a pure Nash equilibrium (if it exists),
recent hardness results regarding
the computation of pure Nash equilibria~\citep{FPT04}
suggest that the best-response dynamics (or any other dynamics)
might take super-polynomial time in the number of players to reach an equilibrium.
Thus, even in case only one (pure) Nash equilibrium exists,
the players might take very long to reach it and thus it
cannot be taken to describe the state of the players
(unless we are willing to ignore the super-polynomially long transient
phase).
In contrast, in the setting studied in this paper these drawbacks
disappear: the solution concept is
{\em defined} in terms of dynamics and for our specific dynamics we have a unique possible prediction.
For rapidly mixing chains the equilibrium is quickly reached.
The concept of metastable distribution and the results in this paper
show that, even for the non-rapidly mixing case,
the system can be often efficiently described after a short initial transient phase.

The Nash Equilibrium concept is based on the assumption that each player
has complete information about the game and the strategies of his opponents
and it is able to compute his best strategy with respect to the strategies
played by the other players.
However, in many complex systems, environmental factors can influence the
way each agent selects her own strategy and limitations to the players'
computational power can influence their behaviors.  Logit dynamics
are a clear and crispy way to model these settings.

\subsection{Related Work}
\paragraph{Logit dynamics.}
Logit dynamics were first  studied by
\citet{blumeGEB93}
who showed that, for two-player two-strategy coordination games, the long-term behavior
of the system is concentrated in the risk dominant equilibrium
(see \citep{hsMIT88}).
\citet{ellisonECO93} initiated the analysis of
how social networks affect
the spread of innovations.
Ellison considered dynamics that are slightly different
from the logit dynamics and
he focused on two extremal network topologies,
the clique and the ring
(the classes of games we study in
Section~\ref{sec::coordmeta}).
His results show
that some large fraction of the players will eventually
choose the strategy with maximum potential.
Similar results were obtained by
\citet{youngTR00}
for the logit dynamics and more general families of graphs.
\citet{msFOCS09}
gave bounds, in terms of some graph theoretic properties of the
underlying interaction network,
on the time the logit dynamics take to hit
the highest potential equilibrium.
\citet{asWINE09}
studied the hitting time of the
Nash equilibrium for the logit dynamics in a class of congestion games.
Bounds on the convergence to specific equilibria have been given also in \cite{kreindler2013fast,kreindler2014rapid}.

The study of the mixing time of the logit dynamics for strategic games has been initiated in
\citep{afppSAGT10} (see also~\citep{afpppSPAA11}), whose results highlight a
separation between games where the mixing time can be bounded independently
from the parameter $\beta$ and games where the mixing time is
necessarily exponential in $\beta$.

Several variants of these dynamics have been proposed
for evaluating the robustness of known results \cite{marden2010revisiting,alos2010logit,alos2015robust,auletta2015logit},
for adapting them to different setting \cite{shah2010dynamics},
or for speeding up their convergence to equilibria \cite{borowski2013fast}.

\paragraph{Metastability.}
The goal of metastability is to model processes showing the following
typical behavior: starting from a given \emph{state},
the system rather quickly visits a nearby
\emph{metastable state} and then it remains very close to such a
state for a very long time;
at some point, the system leaves the metastable state (and its neighborhood) and moves to some other metastable state;
the process then is repeated.
\emph{Metastable states} are identified with subsets of profiles of the system,
with the property that the probability to leave these subsets is exponentially small \cite{begkCMP02}.
Similar concepts have been also introduced in the context of population protocols under the name of \emph{locally stable states}
(see, e.g., \cite{sandholmMIT10} and references therein).
Usually, these states correspond to ``deep'' local energy minima
(or local maximizers of the potential function, in the case of potential games),
with absolute energy minima corresponding to stable states.
Sometimes, metastable state are described through probability distributions
over the profiles in the subset \cite{begkCMP02, bgArxiv11}.
Anyway, the metastability property always refers to the subset and not to the distribution.
We depart from this approach by considering metastable {\em distributions}
and not just metastable subsets of profiles.
Indeed this is necessary for our object of study (logit dynamics of strategic games)
as we shall show that there exist games that admit quite natural metastable distributions
even if in the support there is no strict local maximizers of the potential function (see Section~\ref{sec::ormeta}).

The classical approach to metastability is concerned with evaluating
the time the system takes to escape from a metastable state \cite{keiSpring79}.
More recent approaches \citep{fwSPRINGER84} based on large deviation theory try to describe
also the typical trajectory that a system takes, that is the sequence of states bringing to
the stable one.
This approach has been recently applied also to the logit dynamics \cite{THEC:THEC187},
under the hypothesis that both $\beta$ and $n$ tend to $\infty$.
Better result on the escape time
have been achieved recently, through the so-called potential theoretical approach \cite{begkPTRF01},
that links hitting time theory with spectral properties of the transition matrix of the dynamics.
In contrast our work
not only tries to bound the time the system takes to escape from a metastable distribution,
but also the time the system takes to enter this distribution from a specific
starting profile.

Very recently, and independently from our work,
Friedlin and Koralov \cite{Freidlin2017} introduced a notion of a metastable distribution that resembles the one presented in this work.
Unlike ours, the notion of metastable distribution in \cite{Freidlin2017} does not require the chain to stay close to this distribution for a given number of steps.
Moreover, in \cite{Freidlin2017} it is not considered the time required to get close to a metastable distribution.

\paragraph{Quasi-stationarity and Censored dynamics.}
A \emph{quasi-stationary distribution} \cite{quasi-stationary} describes the limiting behavior of a system
in a subset of states given that the system never leaves this subset.
Quasi-stationary distributions are often associated with metastability (see, e.g., \cite{bgArxiv11}).
Anyway, we highlight that, even if quasi-stationary distributions can be metastable,
the inverse is not necessarily true.
Moreover, works about quasi-stationary distributions focus on bounding the time
the dynamics take to converge to this distribution from the support.
In this work we are also interested in the convergence time to metastable distributions from
profiles that are not in the support.

The work on censored Glauber dynamics \citep{llpPTRF10,dlpJSP09,dlpCMP09} is
also related to ours: the mixing time in censored dynamics resembles
the pseudo mixing time for the metastable distribution on a subset of
states. However, we stress that the censored dynamics alter the
original evolution of the Markov chain
and the techniques developed do not seem useful to answer
questions about the pseudo-mixing time.

\subsection{Notations}
We write $|S|$ for the size of a set $S$.
We use bold symbols for vectors and the standard game theoretic
notation $(\x_{-i},y)$ to denote the vector obtained from $\x$
by replacing its $i$-th entry with $y$;
i.e., $(\x_{-i},y)=(x_1,\ldots,x_{i-1},y,x_{i+1},\ldots,x_n)$.
When we are given a probability distribution $\mu_x$ that assigns probability $1$ to state $x$,
we say that $\mu_x$ is \emph{concentrated} in state $x$.
We refer the reader to Appendix~\ref{apx::mcsummary} for a review
of useful facts about the concepts of \emph{total variation distance}
$\tv{\mu - \nu}$ between probability distributions,
and of \emph{mixing time} $\tm$.

\section{Logit dynamics}\label{sec::logit}
A \emph{strategic game} $\mathcal{G}$ is a triple $\mathcal{G} = ([n], \mathcal{S}, \mathcal{U})$.
The set $[n]=\{1,\dots,n\}$ is a finite set of \emph{players};
$\mathcal{S}=S_1\times\dots\times S_n$ is the set of
{\em strategy profiles} and the finite set $S_i$
is the set of \emph{strategies} for player $i$;
$\mathcal{U}=(u_1,\dots,u_n)$ is a tuple of \emph{utility functions} (or \emph{payoffs})
where $u_i\colon\mathcal{S}\rightarrow\mathbb{R}$ and
$u_i(\x)$ is the payoff of player $i$ in strategy profile $\x$.

{\em Dynamics} for a strategic game are probabilistic rules by which at each step:
(i) one or more players are selected; (ii) each selected player updates her strategy by sampling according
to a probability distribution that may depend on the current strategy profile.
For example, in the best response dynamics each selected player updates her strategy by selecting
with probability 1
a strategy that maximizes her payoff given the strategies currently adopted by other players.

In this paper we study specific dynamics for strategic games,
the {logit dynamics}, introduced by Blume~\citep{blumeGEB93} and referred to as the
{\em log-linear model}.
The \emph{logit dynamics with parameter $\beta\geq 0$} are randomized best-response dynamics
where at each step one player $i\in[n]$ is selected uniformly at random and she updates her strategy
in profile $\x\in\mathcal{S}$ by choosing strategy $y\in S_i$ with probability $\sigma_i(y\mid\x)$
defined as follows
\begin{equation}\label{eq:updateprob}
\sigma_i(y \mid \x) = \frac{e^{\beta u_i(\x_{-i}, y)}}{T_i(\x)},
\end{equation}
where
$T_i(\x)=\sum_{z \in S_i} e^{\beta u_i(\x_{-i}, z)}$ is the normalizing factor.
In other words, in the logit dynamics the logarithm of the ratio of the probability of
choosing strategy $y$ and $y'$ is proportionally related through $\beta$ to the difference of
the respective utilities.
Parameter $\beta$ can be seen as descriptive of the \emph{rationality level} of the players:
for $\beta = 0$ (no rationality) player $i$ updates
her strategy by selecting a new strategy uniformly at random;
for $\beta>0$, the probability is biased towards strategies yielding higher payoffs;
and for $\beta\rightarrow\infty$ (full rationality) player $i$ chooses her best response strategy
(if more than one best response is available, she chooses uniformly at random one of them).

For every $\beta \geq 0$, the logit dynamics with parameter $\beta$ induce a Markov chain
over the set $\mathcal{S}$ of strategy profiles.
More formally, we have the following definition.
\begin{definition}[Markov chain of the logit dynamics~\citep{blumeGEB93}]
Let $\mathcal{G} = ([n], \mathcal{S}, \mathcal{U})$ be a strategic game and let $\beta \geq 0$.
The \emph{Markov chain $\mathcal{M}_\beta$ of the logit dynamics with parameter $\beta$}
for strategic game $\mathcal{G}=([n],\mathcal{S},\mathcal{U})$ is the Markov chain
with state space $\mathcal{S}$ and transition probability
\begin{equation}\label{eq:transmatrix}
P(\x, \y) = \frac{1}{n} \cdot
\begin{cases}
\sum_{i=1}^n \sigma_i(y_i \mid \y), & \mbox{ if } \x = \y; \\
\sigma_i(y_i \mid \x), & \mbox{ if } \x\ne\y \mbox{ and } \y_{-i} = \x_{-i}; \\
0, & \mbox{ otherwise;}
\end{cases}
\end{equation}
where $\sigma_i(y_i \mid\x)$ is defined in \eqref{eq:updateprob}.
\end{definition}
It is easy to see that the Markov chain is ergodic as $P^n(\x, \y) > 0$ for every pair of profiles $\x$ and $\y$. Hence, a unique stationary distribution, i.e. a distribution $\pi$ such that $\pi = \pi P$, exists and for every starting profile $\x$
the distribution $P^t(\x, \cdot)$ approaches $\pi$ as $t$ tends to infinity.

\paragraph{Potential games.}
Function $\Phi\colon \mathcal{S}\rightarrow \mathbb{R}$ is a {\em potential} for strategic game
$\mathcal{G}=([n],\mathcal{S},\mathcal{U})$ if
for every player $i\in[n]$ and for profiles $\x,\y\in\mathcal{S}$ differing only for player $i$,
it holds that $u_i(\x)-u_i(\y)=\Phi(\x)-\Phi(\y)$.
Games $\mathcal{G}$ admitting a potential function are called \emph{potential games}~\citep{MS96}.
The stationary distribution of the Markov chain of the logit dynamics with parameter $\beta$ of a
potential game $\mathcal{G}$ is the Gibbs measure $\pi$ defined as follows
\begin{equation}\label{eq:Gibbs}
\pi(\x) = \frac{e^{\beta \Phi(\x)}}{Z},
\end{equation}
where
$Z=\sum_{\y \in \mathcal{S}} e^{\beta\Phi(\y)}$ is the {\em partition function}.
The Markov chain of the logit dynamics with parameter $\beta$ for a potential game $\mathcal{G}$ coincides
with the well-studied \emph{Glauber dynamics} with temperature $1/\beta$ for the Gibbs measure $\pi$.

\paragraph{Two-strategy coordination games.}
A two-strategy coordination game is a two-player strategic game where each player has
two strategies $+1$ and $-1$. Thus $S_1=S_2=\{\pm 1\}$ and
$\mathcal{S}=\{\pm 1\}^2$. The utility function  is described by the following table
$$
\begin{game}{2}{2}
      &$+1$      &$-1$ \\
  $+1$ &$a,a$    &$c,d$ \\
  $-1$ &$d,c$    &$b,b$
\end{game}\hspace*{\fill}
$$
Thus, for example, $u_1(-1,+1)=d$ and $u_2(-1,+1)=c$.
It is usually assumed that $a>d$ and $b>c$ to model the fact that each player prefers
to choose the same strategy as the other player
(hence the name of {\em coordination} games).
This implies that profiles
$(+1,+1)$ and $(-1,-1)$ are Nash equilibria of the two-strategy coordination games.
We define $\Delta:=a-d$ and $\delta:=b-c$ and assume, w.l.o.g., that $\Delta \geq \delta$.
If $\Delta>\delta$, then $+1$ is the
{\em risk dominant strategy} (\cite{hsMIT88}).
It is easy to see that the function $\Phi$ defined as follows
$$
 \Phi(+1,+1)=\Delta,\quad\Phi(-1,-1)=\delta,\quad\mbox{\rm and \ } \Phi(+1,-1)=\Phi(-1,+1)=0
$$
is a potential function for the two-strategy coordination game.
The logit dynamics with parameter $\beta$ for two-strategy coordination games have been first studied in
\cite{blumeGEB93} and have the following transition matrix
$$
P =
\left(
\begin{array}{c|cccc}
 & (+1,+1) & (+1,-1) & (-1,+1) & (-1,-1) \\
\hline
(+1,+1) & 1 - p & p/2 & p/2 & 0 \\[1mm]
(+1,-1) & (1- p)/2 & (p + q)/2 & 0 & (1- q)/2 \\[1mm]
(-1,+1) & (1-p)/2 & 0 & (p + q)/2 & (1 - q)/2 \\[1mm]
(-1,-1) & 0 & q/2 & q/2 & 1 - q
\end{array}
\right)
$$
where
$p = \frac{1}{1+e^{\Delta\beta}}$ and $q = \frac{1}{1+e^{\delta\beta}}$.
The partition function is
$$Z=e^{\beta\Delta}+1+1+e^{\beta\delta}=\frac{1}{p}+\frac{1}{q}$$
and the stationary distribution for $P$ is the Gibbs measure
$$
\pi = \left(\frac{e^{\beta\Delta}}{Z},\frac{1}{Z},\frac{1}{Z},\frac{e^{\beta\delta}}{Z}\right)$$
that can be re-written as
$$
\pi = \left(\frac{q(1-p)}{p + q}, \frac{pq}{p + q}, \frac{pq}{p + q}, \frac{p(1-q)}{p + q}\right).
$$

\section{Metastability}\label{sec::meta}
In this section we give formal definitions of the concepts of
\emph{metastable distribution} (Section~\ref{subsec:meta}) and
of \emph{pseudo-mixing time} (Section~\ref{subsec:pmt}) and we highlight
some of their properties.
% In Section~\ref{sec::bottleneck}, we highlight connections between metastability and the bottleneck ratio.
% In Section~\ref{subsec:tools}, we instead highlight connections between pseudo-mixing and hitting time.
Finally, we exemplify the new concepts by studying a simple three-state Markov chain and
the Markov chain of the logit dynamics for two-player two-strategy coordination games
(in Section~\ref{sec::ormeta} we will study an $n$-player pure coordination game
and in Section~\ref{sec::graphmeta} we study the metastability of coordination games on graphs).

\subsection{Metastable distributions}
\label{subsec:meta}
\begin{definition}[Metastable distribution]\label{def:metastability}
Let $P$ be a Markov chain with finite state space $\Omega$.
A probability distribution $\mu$ over $\Omega$ is $(\varepsilon,T)$\emph{-metastable} for
$P$ if, for $0 \leq t \leq T$, it holds that
$$
\tv{\mu P^t - \mu} \leq \varepsilon.
$$
\end{definition}
The definition of a metastable distribution captures the idea of a distribution that behaves approximately
like the stationary distribution; meaning that if we start from a metastable distribution
and run the chain we stay close to that distribution (that is, within $\epsilon$) for a long time
(that is, for at least $T$ time steps).

\medskip Let us now highlight some properties of metastable distributions, that can be easily derived from known results in Markov chain theory:
\begin{enumerate}
\item \emph{Monotonicity}: If $\mu$ is $(\varepsilon,T)$-metastable for $P$
then it is $(\varepsilon',T')$-metastable for every $\varepsilon' \geq \varepsilon$ and
$T' \leq T$;
\item \emph{Stationarity}:
$\mu$ is stationary if and only if it is $(0,1)$-metastable.
\end{enumerate}
A third property is given by the following easy and useful lemma.
\begin{lemma}[\emph{Additivity}]
\label{lem:meta:1}
If $\mu$ is
{$(\varepsilon,1)$-metastable} for $P$ then, for every integer $T>0$,
 $\mu$ is {$(\varepsilon T,T)$-metastable} for $P$.
\end{lemma}
\begin{proof}
Since the total variation distance satisfies the triangle inequality (see Fact~\ref{fact:triangleTV} in Appendix~\ref{apx::mcsummary}) and $\mu$ is $(\varepsilon, 1)$-metastable, we have
$$
 \tv{\mu P^T - \mu} \leq \tv{\mu P^T - \mu P} + \tv{\mu P - \mu} \leq \tv{\mu P^{T-1} - \mu} + \varepsilon.
$$
The lemma then follows.
\end{proof}
The next lemma states that the convex combination of two metastable distributions is metastable.
\begin{lemma}[\emph{Convexity}]
\label{lem:comb}
If $\mu_1$ is $(\varepsilon_1,T_1)$-metastable for $P$ and
   $\mu_2$ is $(\varepsilon_2,T_2)$-metastable for $P$,
then, for $0\leq\alpha\leq 1$, $\mu = \alpha \mu_1 + (1-\alpha) \mu_2$
is $(\varepsilon, T)$-metastable for $P$ with
$\varepsilon = \max \{\varepsilon_1, \varepsilon_2\}$ and $T = \min \{T_1, T_2\}$.
\end{lemma}
\begin{proof}
 For any $t \leq T$, we have
 \begin{align*}
  \tv{\mu P^t - \mu} & = \tv{\alpha \left(\mu_1 P^t - \mu_1\right) + (1-\alpha) \left(\mu_2 P^t - \mu_2\right)}\\
  & \leq \alpha \tv{\mu_1 P^t - \mu_1} + (1-\alpha) \tv{\mu_2 P^t - \mu_2} \leq \varepsilon.\qedhere
 \end{align*}
\end{proof}
Finally, we highlight a connection between metastability
and \emph{bottleneck ratio}.
Given an ergodic Markov chain $P$ with state space $\Omega$ and stationary distribution $\pi$,
the bottleneck ratio $\Bn(S)$ for a subset $S\subseteq \Omega$ of states, is defined as
$$
\Bn(S) = \frac{Q(S,\overline{S})}{\pi(S)},
$$
where $Q(S,\overline{S}) = \sum_{x\in S}\sum_{y\in\Omega\setminus S} \pi(x) P(x,y)$.
Let $\pi_S$ be the stationary distribution conditioned on $S$; i.e.,
\begin{equation}
\label{eq:piS}
\pi_S(x)=\begin{cases}
\pi(x)/\pi(S),& \text{if } x\in S;\cr
0,& \text{otherwise}.
\end{cases}
\end{equation}
\begin{lemma}
\label{lem:meta:bottleneck}
Let $P$ be a Markov chain with finite state space $\Omega$ and let
$S\subseteq\Omega$ be a subset of states. Then, $\pi_S$ is $(\Bn(S),1)$-metastable.
\end{lemma}
\begin{proof}
The lemma follows from the fact that
the bottleneck ratio $\Bn(S)$ of
$S$ is equal to the total variation distance between $\pi_S$ and $\pi_S P$;
i.e., $\|\pi_S P - \pi_S \| = B(S)$.
See, for example, Theorem~7.3 from~\citep{lpwAMS08}).
\end{proof}
%%PINO

\subsection{Pseudo-mixing time}
\label{subsec:pmt}
Among all metastable distributions,
we are interested in the ones that are quickly reached from a, possibly large, set of states.
This motivates the following definition.

\begin{definition}[Pseudo-mixing time]\label{def:pseudomixing}
Let $P$ be a Markov chain with state space $\Omega$,
let $S \subseteq \Omega$ be a set of states and
let $\mu$ be a probability distribution over $\Omega$.
We define $d_\mu^S(t)$  as
$$
 d_\mu^S(t) = \max_{x\in S} \tv{P^t(x,\cdot)-\mu}.
$$
Then the \emph{pseudo-mixing time} $\pmt{\mu}{S}{\epsilon}$ of $\mu$ starting from $S$ is
$$
\pmt{\mu}{S}{\varepsilon} = \inf \left\{ t \in \mathbb{N} \colon d_\mu^S(t) \leq \varepsilon \right\}.
$$
\end{definition}
The pseudo-mixing time extends the concept of mixing time to metastable distributions and
it coincides with the mixing time for a stationary distribution.
Indeed, the stationary distribution $\pi$ of an ergodic Markov chain is
a $(0,1)$-metastable distribution that is is reached within $\varepsilon$
in time $\tm(\varepsilon)$ from every state (see Appendix~\ref{apx::mcsummary}).
Thus,
according to Definition~\ref{def:pseudomixing}, we have that
$\pmt{\pi}{\Omega}{\varepsilon}=\tm(\varepsilon)$.

\medskip We now highlight some interesting properties related to the pseudo-mixing time and metastable distribution concepts.
The first property connects $d_\mu^S(t)$ for a metastable distribution $\mu$
to a quantity that does not depend on this distribution.
% The quantity $d_\mu^S(t)$ is related to the pseudo-mixing time $\pmt{\mu}{S}{\epsilon}$ by the
% following simple observation.
% \begin{obs}
% \label{obs:pmt}
% Let $\mu$ be a metastable distribution for Markov chain $P$ with space set $\Omega$ and
% let $S\subseteq\Omega$.
% If, for $t>T$, $d_\mu^S(t)\leq\epsilon$ then $\pmt{\mu}{S}{\epsilon}\leq T$.
% \end{obs}
% We next show two interesting properties of the pseudo-mixing time.
\begin{lemma}
\label{lem:pmt:coupling}
Let $P$ be a Markov chain with finite state space $\Omega$ and let
$\mu$ be an $(\epsilon,T)$-metastable distribution supported over a subset
$S$ of the state space.
Then, for  $1\leq t \leq T$, it holds that
$$
 d_\mu^S(t)\leq \epsilon + \max_{x,y\in S} \tv{P^t(x,\cdot)-P^t(y,\cdot)}.
$$
\end{lemma}
\begin{proof}
From the triangle inequality, we have
$$
\tv{P^t(x, \cdot) - \mu} \leq \tv{P^t(x, \cdot) - \mu P^t} + \tv{\mu P^t - \mu}.
$$
Since $\mu$ is $(\varepsilon, t)$-metastable, then for every $t \leq T$ we have
$\tv{\mu P^t - \mu} \leq \varepsilon$.
Moreover, since $\mu(y)=0$ for $y \notin S$, for every set of states $A\subseteq\Omega$ and for every $t$ it holds that
\begin{align*}
|P^t(x, A) - \mu P^t(A)| & = \left| P^t(x, A) - \sum_{y \in S} \mu(y) P^t(y, A) \right|
= \left| \sum_{y \in S} \mu(y) \left( P^t(x, A) - P^t(y, A) \right) \right| \\
& \leq \sum_{y \in S} \mu(y) \left| P^t(x, A) - P^t(y, A) \right|
\leq \max_{y \in S} \left| P^t(x, A) - P^t(y, A) \right|.
\end{align*}
Thus, the total variation between $P^t(x,\cdot)$ and $\mu P^t$ is
\begin{align*}
\tv{P^t(x,\cdot) - \mu P^t} & = \max_{A \subseteq \Omega} |P^t(x, A) - \mu P^t(A)| \\
& \leq \max_{A \subseteq \Omega} \max_{y \in S} \left| P^t(x, A) - P^t(y, A) \right| = \max_{y \in S} \tv{P^t(x, \cdot) - P^t(y, \cdot)}. \qedhere
\end{align*}
\end{proof}
Notice that if $\mu$ is the stationary distribution then it is $(0,T)$-metastable for every $T$ and the above lemma gives
$$
d_\mu^\Omega(t)\leq \max_{x,y\in \Omega} \tv{P^t(x,\cdot)-P^t(y,\cdot)}
$$
for all $t$.
This is a well-know inequality and it is widely used to bound the mixing
time of Markov  chains (see, for example, Lemma~4.11 in~\citep{lpwAMS08}).

Observe that, even if the pseudo-mixing time of a metastable distribution
$\mu$ is finite, we cannot argue that the Markov chain converges to $\mu$,
but only that it converges to a distribution $\mu'$ that is \emph{close} to $\mu$.
However, next lemma shows that when starting from $\mu'$ the Markov chain
will behave approximatively as if it starts from the metastable distribution $\mu$.
\begin{lemma}\label{lem:metaonetot}
Let $P$ be a Markov chain with finite state space $\Omega$ and let
$\mu$ be an $(\epsilon,T)$-metastable distribution supported over a subset
$S$ of the state space with  $t_\mu^S(\varepsilon) < + \infty$.
Then, for every $x\in S$ and
$t_\mu^S(\varepsilon)\leq t\leq t_\mu^S(\varepsilon) + T$,
it holds that
$$
\tv{P^t(x,\cdot) - \mu} \leq 2 \varepsilon .$$
\end{lemma}
\begin{proof}
Let us name $\bar{t} = t - t_\mu^S(\varepsilon)$ for convenience sake. By using the triangle inequality for the total variation distance, the fact that $P^{\bar{t}}$ is a stochastic matrix, and the definitions of metastable distribution and pseudo-mixing, we have that
\begin{align*}
\tv{P^t(x,\cdot) - \mu} & = \tv{P^{t_\mu^S(\varepsilon)}(x,\cdot) P^{\bar{t}} - \mu} \\
& \leq \tv{P^{t_\mu^S(\varepsilon)}(x,\cdot)  P^{\bar{t}} - \mu P^{\bar{t}}} + \tv{\mu P^{\bar{t}} - \mu} \\
& \leq \tv{P^{t_\mu^S(\varepsilon)}(x,\cdot) - \mu} + \tv{\mu P^{\bar{t}} - \mu} \leq 2 \varepsilon. \qedhere
\end{align*}
\end{proof}

Finally, we highlight some connections between pseudo-mixing time and \emph{hitting time}.
Given a Markov chain $P$ with finite state space $\Omega$,
the hitting time $\tau_S$ of a subset $S \subset \Omega$ is
the first time step in which the Markov chain reaches a state from $S$, i.e.,
$$
 \tau_S = \min \{ t \colon X_t \in S\}.
$$
When $S$ consists of a single state $x$, we write $\tau_x$ rather then $\tau_{\{x\}}$.
Then, we have the following lemma.
\begin{lemma}
\label{lem:pmt:hitting}
Let $P$ be a Markov chain with finite state space $\Omega$ and let
$\mu_y$ be an $(\epsilon,T)$-metastable distribution concentrated in state $y$.
Then for all $S\subseteq\Omega$ and $t \leq T$  we have
$$
d_{\mu_y}^S(t)\leq \epsilon +\max_{x\in S} \Prob{x}{\tau_y > t}.
$$
\end{lemma}
\begin{proof}
Since $\mu_y$ is concentrated in $y$, we have that
\begin{align*}
\tv{P^t(x,\cdot)-\mu_y} = \Prob{x}{X_t \neq y} &= \Prob{x}{X_t \neq y, \tau_y \leq t} + \Prob{x}{X_t \neq y, \tau_y> t}\\
& = \Prob{x}{X_t \neq y \mid \tau_y \leq t}\Prob{x}{\tau_y \leq t} + \Prob{x}{\tau_y > t}.
\end{align*}
Moreover, observe that
\begin{align*}
\Prob{x}{X_t \neq y \mid \tau_y \leq t} & = \sum_{k \leq t} \Prob{x}{X_t \neq y \mid \tau_y = k}\Prob{x}{\tau_y = k \mid \tau_y \leq t} \\
& = \sum_{k \leq t} \Prob{y}{X_{t-k} \neq y}\Prob{x}{\tau_y = k \mid \tau_y \leq t} \\
& = \sum_{k \leq t} \tv{\mu_y P^{t-k} - \mu_y} \Prob{x}{\tau_y = k \mid \tau_y \leq t}\\
& \leq \varepsilon \sum_{k \leq t}\Prob{x}{\tau_y = k \mid \tau_y \leq t} = \varepsilon.
\end{align*}
where in the inequality we used the metastability of $\mu_y$. Hence,
\begin{align*}
\tv{P^t(x,\cdot)-\mu_y} & = \Prob{x}{X_t \neq y \mid \tau_y \leq t}\Prob{x}{\tau_y \leq t} + \Prob{x}{\tau_y > t} \\
& \leq \varepsilon \Prob{x}{\tau_y \leq t} + \Prob{x}{\tau_y > t} \leq \varepsilon + \Prob{x}{\tau_y > t}.\qedhere
\end{align*}
\end{proof}
The following lemma show another interesting relation between hitting time and the pseudo-mixing time of a convex combination of metastable distributions.
\begin{lemma}
 \label{lem:hitting_convex}
 Let $P$ be a Markov chain with finite state space $\Omega$.
 Let $\mu_y$ be an $(\epsilon_y,T_y)$-metastable distribution concentrated in state $y$
 and $\mu_z$ be an $(\epsilon_z,T_z)$-metastable distribution concentrated in state $z$.
 Fix $\varepsilon = \max \{\varepsilon_y, \varepsilon_z\}$ and $T = \min \{T_y, T_z\}$.
Then for all $x \in \Omega$ and $t \leq T$ we have
$$
d_{\mu_{x,t}}^{\{x\}}(t) \leq \epsilon + \Prob{x}{\tau_{\{y,z\}} > t},
$$
where $\mu_{x,t} = \alpha_{x,t} \mu_y + (1 - \alpha_{x,t}) \mu_z$ with $\alpha_{x,t} = \Prob{x}{\tau_y \leq \tau_z \mid \tau_{\{y,z\}} \leq t}$.
\end{lemma}
\begin{proof}
 For sake of readability, let us denote by $E$ the event $\tau_{\{y,z\}} > t$ and by $\overline{E}$ its complement. Observe that
\begin{align*}
 \tv{P^{t}(x,\cdot) - \mu_{x,t}} & = \max_{A \subset \Omega} \left| \Prob{x}{X_{t} \in A} - \mu_{x,t}(A)\right|\\
 & = \max_{A \subset \Omega} \left| \Prob{x}{X_{t} \in A \mid \overline{E}} - \mu_{x,t}(A) + \left(\Prob{x}{X_{t} \in A \mid E} - \Prob{x}{X_{t} \in A \mid \overline{E}}\right) \Prob{x}{E}\right|\\
 & \leq \Prob{x}{E} + \max_{A \subset \Omega} \left| \Prob{x}{X_{t} \in A \mid \overline{E}} - \mu_{x,t}(A)\right|.
\end{align*}
Since Markov chains are memoryless, we have
$$
 \Prob{x}{X_{t} \in A \mid \overline{E}} = \sum_{w} \Prob{x}{X_{\tau_{\{y,z\}}} = w \mid \overline{E}} \Prob{x}{X_{t} \in A \mid X_{\tau_{\{y,z\}}} = w \wedge \overline{E}}.
$$
Observe that
$\Prob{x}{X_{\tau_{\{y,z\}}} = y \mid \overline{E}} = \alpha_{x,t}$,
$\Prob{x}{X_{\tau_{\{y,z\}}} = z \mid \overline{E}} = 1 - \alpha_{x,t}$,
and $\Prob{x}{X_{\tau_{\{y,z\}}} = w \mid \overline{E}} = 0$ for each state $w \neq y,z$.
That is, $\Prob{x}{X_{\tau_{\{y,z\}}} = w \mid \overline{E}} = \mu_{x,t}(w)$ for each $w \in \Omega$.
Moreover, since $t - \tau_{\{y,z\}} \geq 0$ if the event $\overline{E}$ holds, for each $w \in \Omega$ we have
$$
 \Prob{x}{X_{t} \in A \mid X_{\tau_{\{y,z\}}} = w \wedge \overline{E}} = \Prob{w}{X_{t - \tau_{\{y,z\}}} \in A} = P^{t - \tau_{\{y,z\}}}(w, A).
$$
Thus,
$$
 \tv{P^{t}(x,\cdot) - \mu_{x,t}} \leq \Prob{x}{E} + \max_{A \subset \Omega} \left| \left(\mu_{x,t}P^{t - \tau_{\{y,z\}}}\right)(A) - \mu_{x,t}(A)\right| \leq \Prob{x}{E} + \varepsilon,
$$
where the last inequality follows from $\mu_{x,t}$ being $(\varepsilon,T)$-metastable from Lemma~\ref{lem:comb} and since $t - \tau_{\{y,z\}} \leq T$.
\end{proof}

\subsection{Examples}
\subsubsection{A simple three-state Markov chain}
As a first example,
let us consider the simplest Markov chain that highlights
the concepts of \emph{metastability} and \emph{pseudo-mixing},

\begin{minipage}{8cm}
$$
P =
\left(
\begin{array}{ccc}
\varepsilon & \frac{1-\varepsilon}{2} & \frac{1-\varepsilon}{2} \\
\varepsilon & 1 - \varepsilon & 0 \\
\varepsilon & 0 & 1 - \varepsilon
\end{array}
\right)
$$
\end{minipage}
\begin{minipage}{5cm}
 \begin{tikzpicture}[->,>=stealth,shorten >=1pt,auto,node distance=1cm,on grid=true,semithick,
                     prof/.style={shape=circle,fill=gray,draw,text=white,inner sep=0pt,minimum size=6mm},
                     every label/.style={font=\scriptsize}]
  \node[prof] (A) {0} edge [loop above] node {$\varepsilon$} ();
  \node[prof] (B) [below left=1.5cm of A] {1} edge [loop below] node {$1 - \varepsilon$} ();
  \node[prof] (C) [below right=1.5cm of A] {2} edge [loop below] node {$1 - \varepsilon$} ();

  \draw [->] (A.west) -- node[swap] {$\frac{1 - \varepsilon}{2}$} (B.north);
  \draw [->] (B.north east) -- node[swap] {$\varepsilon$} (A.south west);
  \draw [->] (A.east) -- node {$\frac{1 - \varepsilon}{2}$} (C.north);
  \draw [->] (C.north west) -- node {$\varepsilon$} (A.south east);
\end{tikzpicture}
\end{minipage}
\par\noindent
The chain is ergodic with stationary distribution
$\pi = \left( \varepsilon, (1 - \varepsilon)/2, (1 - \varepsilon)/2 \right)$ and its mixing time
is $\tm = \Theta\left(1/\varepsilon\right)$. Hence the mixing time grows unbounded
as $\varepsilon$ tends to zero.

Now observe that, for every $\delta > \varepsilon$, distributions
$\mu_1 = (0,1,0)$ and $\mu_2 = (0,0,1)$ are $(\delta, \Theta(\delta/\varepsilon))$-metastable
according to Definition~\ref{def:metastability}.
Moreover, if the chain starts from state $0$,
the state of the chain after one step is distributed as in the stationary distribution.
Hence,
even if the mixing time can be arbitrary large,
for every $\epsilon$ and for every starting state $x$ there is a $(\delta, \Theta(\tm))$-metastable distribution
$\mu$ that is quickly (in constant, independent of $\varepsilon$, time) reached from $x$.

\subsubsection{Two-strategy coordination games}
In~\citep{afppSAGT10} it is proved that the mixing time of the logit dynamics with
parameter $\beta$ for a two-player two-strategy coordination game defined above is
$\tm=\Theta(1/q)=\Theta(1+e^{\beta\delta})$ and thus it grows unbounded in $\beta$.
We next describe distributions that are metastable for an amount of time of the same order
of the mixing time and whose pseudo mixing time is independent from $\beta$.
Specifically, consider distributions
$\mu_+$ and $\mu_-$ concentrated in states $(+1,+1)$ and $(-1,-1)$, respectively; i.e.,
$$
\mu_+ = \left(1,0,0,0\right) \quad \mbox{ and } \quad \mu_- = \left(0,0,0,1\right).
$$
Observe that, if we start from $\mu_+$ or $\mu_-$, after one step of the chain we are respectively in distributions
$$
\mu_+ P = \left(1-p,  p/2,  p/2,  0\right)
\quad \mbox{ and } \quad
\mu_- P = \left(0,  q/2,  q/2,  1-q\right).
$$
Then, since $\Delta \geq \delta$, we obtain
$$
\tv{\mu_+ P - \mu_+} = p \leq q
\quad \mbox{ and } \quad
\tv{\mu_- P - \mu_-} = q.
$$
In other words, $\mu_-$ and $\mu_+$ are $(q,1)$-metastable for the Markov chain
of the logit dynamics with parameter $\beta$ for the two-strategy coordination game.
By Lemma~\ref{lem:meta:1}, they are
$(\varepsilon, \varepsilon/q)$-metastable for any $\varepsilon > 0$
and, since each of the two distributions has a support set consisting of one element,
the pseudo-mixing time starting from their respective supports is trivially $0$.

Let us discuss what happens when the Markov chain starts from the states
$(+1,-1)$ or $(-1,+1)$.
The distributions after one step starting from the two states are
$$
P((+1,-1), \cdot) = \left( \frac{1-p}{2}, \frac{p+q}{2},  0,  \frac{1-q}{2} \right),
\quad \mbox{ and } \quad
P((-1,+1), \cdot) = \left( \frac{1-p}{2}, 0,  \frac{p+q}{2},  \frac{1-q}{2} \right).
$$
Thus, we compute the total variation distance
from the stationary distribution $\pi$ and obtain
$$
\tv{P((+1,-1), \cdot) - \pi} = \tv{P((-1,+1), \cdot) - \pi} = \frac{p}{2} + \frac{1}{2} \cdot \frac{q - p}{q + p} < \frac{1}{2},
$$
where we used that $p+q<2$.
Then, we obtain that the logit dynamics starting from
$(+1,-1)$ or $(-1,+1)$ are $\varepsilon$-close to the stationary distribution
after $\OO(\log \varepsilon^{-1})$ steps for each $\varepsilon>0$
(see Fact~\ref{fact:tm} in Appendix~\ref{apx::mcsummary}).

We summarize the above discussion in the following theorem.
\begin{theorem}\label{theorem:2peasy}
For every starting profile $\x\in\{\pm 1\}^2$ and $\varepsilon > 0$,
the Markov chain of the logit dynamics with parameter $\beta$ for the two-strategy coordination game admits
a $\left(\varepsilon,\tm\cdot\varepsilon\right)$-metastable distribution
$\mu_{\x}$ with pseudo-mixing time
$t^{\{\x\}}_{\mu_{\x}} = \OO\left(\log \varepsilon^{-1}\right)$ (independent of $\beta$).
\end{theorem}
We would like to remark how the above theorem,
jointly with the result on the mixing time of \cite{afpppSPAA11},
gives a complete picture of the behavior of a system that evolves according to the logit dynamics for the two-strategy coordination game:
given the initial state, metastable distributions describe, with an error at most $\varepsilon$,
the state of the system for every time $t$ between $\log 1/\varepsilon$ and $(1+e^{\beta\delta})\cdot\varepsilon$,
whereas the stationary distribution can be used for describing the system for larger values of $t$.
%
% the state of this system can be described, after an initial transient phase of length at most $(1+e^{\beta\delta})\log{1/\varepsilon}$, by the stationary distribution.
% If $\beta$ is large this might be unsatisfactory as the initial transient phase becomes too large.
% In this case, given the initial state, we can predict, with an error at most $\varepsilon$,
% the future state of the system for every time $t$ between $\log 1/\varepsilon$ and $(1+e^{\beta\delta})\cdot\varepsilon$.\footnote{Actually, we note that
% states with same number of players adopting the $+1$ strategy reach the same metastable distribution. This phenomenon is not accidental and will manifest also in our study of graphical coordination games.}

\section{Metastability of Pure Coordination games}\label{sec::ormeta}
In this section, we study a simple potential game and describe three metastable distributions
for its logit dynamics.
One of these metastable distributions is particularly interesting:
it gives positive probability to states that are not local maximum of the potential function and
that are easy to leave.
This shows that metastable distributions are not necessarily concentrated
on states that are increasingly hard to leave as $\beta$ grows.

For $n\geq 3$,
the \emph{pure coordination game}~\cite{behavior} (also known as \emph{unanimity game}) is an $n$-player game
where players have the same strategy set $A$
and each players is \emph{happy} when all players adopt the same strategy
and {\em unhappy} otherwise.
More formally, the utility $u_i(\x)$ of player $i\in [n]$ in profile $\x$ is
$u_i(\x) = a$, if $\x=(s,\ldots,s)$ for some $s \in A$;
and $u_i(\x)=b<a$, otherwise.
This class of games has several Nash equilibria (every profile except the ones in which a single player adopts a strategy different from the rest of the players)
strategy is an equilibrium) and this makes it very difficult to predict the behavior of the system.
These games are often used for modeling the division of people in teams (e.g., close friends deciding which soccer team to join),
or the choice of firms' location in a space (e.g., linked industries deciding in which area of the country to produce their goods).

In order to keep the analysis of metastability for the logit dynamics as
simple as possible,
we consider a very simple setting in which the strategy set
$A=\{+1,-1\}$ consists of only two strategies and $a = 1$ and $b=0$.
We let $\p$ and $\m$ denote, respectively,
the profile in which all players play $+1$ and
the profile in which all players play $-1$.
This game is a potential game with the following potential function $\Phi$:
$\Phi(\p)=\Phi(\m)=1$ and $\Phi(\x)=0$ for $\x\ne\p,\m$.
The mixing time of its logit dynamics is roughly $e^\beta$,
and hence super-polynomial for $\beta = \omega(\log n)$.
To see this, recall the known relation between bottleneck ratio and mixing time
(see, e.g., Chapter 7.2 in~\cite{lpwAMS08}) and
observe that the probability of $\p$ at stationarity is $\pi(\p) \leq 1/2$
and the bottleneck ratio for $\p$ is $B(\p) = \frac{1}{1 + e^{\beta}}$.

Let us now focus on the metastable distributions.
We show that if we start the logit dynamics at a profile where at least one player is playing $+1$
and at least one is playing $-1$, then after $\OO(\log n)$ time steps the distribution
is close to uniform on the profiles different from $\p$ and $\m$,
and it stays close to this distribution for exponential time.
Hence, even if there is no small sub-set of the state space where the chain stays close for a long time,
we can still say that the chain is metastable  in the sense that the ``distribution'' of the chain stays
close to some well-defined distribution for a long time.

Specifically, we identify three metastable distributions for the logit dynamics:
$\pi_{\+}$, the distribution concentrated on the profile $\p$ where every player plays $+1$;
$\pi_{-}$,  the distribution concentrated on the profile $\p$ where every player plays $-1$;
and the uniform distribution $U$ over all remaining states.
In addition, we prove that the pseudo-mixing time of $U$ starting from
a profile other than $\p$ and $\m$ is $O(n\log n)$.

The idea behind this result is that the process according to which the number of players adopting a given strategy evolves
is essentially an Ehrenfest urn (see Appendix~\ref{app:ehrenfest}), the only difference being at extreme values,
i.e., when this number is either $0$ or $n$. Hence, as long as one of these extreme value is not hit (and this takes time that is exponential in $n$),
the distribution on the profiles of the game is well-described by the stationary distribution of the Ehrenfest urn as an approximation of the future status.

The next lemma establishes the metastability of $\pi_{\+}$, $\pi_{-}$ and $U$.
\begin{lemma}
\label{lemma:ORuniformmeta}
Consider the logit dynamics with parameter $\beta$ for the $n$-player pure equilibrium game.
For every $\varepsilon > 0$, the uniform distribution $U$ over $R=\{+1,-1\}^n \setminus \{\p, \m\}$ is
$(\varepsilon, \varepsilon\cdot 2^n)$-metastable and
$\pi_{\+}$ and $\pi_{-}$ are $(\varepsilon, \varepsilon \cdot e^{\beta})$-metastable.
\end{lemma}
\begin{proof}
Let us denote by $P$ the transition matrix of the logit dynamics and  by $\pi$ its
stationary distribution. Then observe that $\pi(R)=\frac{2^n-2}{Z}$ and
$Z=2e^{\beta} + 2^n-2$.
Moreover,
$$
Q(R,\overline{R}) = \sum_{\x \in R}\pi(\x)\cdot\bigl(P(\x,\p)+P(\x,\m)\bigr)=
    \frac{2}{Z}\cdot\frac{e^{\beta}}{1+e^\beta}.
$$
Therefore, the bottleneck ratio is
$$
B(R) = \frac{Q(R,\overline{R})}{\pi(R)} = \frac{e^\beta}{1+e^\beta}\cdot\frac{2}{2^n-2} <\frac{1}{2^{n - 1}}.
$$
Similarly, it is immediate to see that $B(\p) = B(\m) = \frac{1}{1 + e^{\beta}}$.
The lemma then follows from Lemma~\ref{lem:meta:bottleneck}, Lemma~\ref{lem:meta:1} and by observing that the stationary distribution restricted to $R$ is the uniform distribution $U$.
\end{proof}
As for bounding the convergence to these metastable distributions, we first start with the following technical lemma.

\newcommand{\hittingorStm}{Let $P$ be the transition matrix of the logit dynamics for the $n$-player OR-game and let $\tau_{\p,\m}$ denote the hitting time of either profile $\p$ or profile $\m$. Then, for every profile $\x$ with at least one player adopting action $+1$ and at least one player adopting action $-1$,
$$
\Prob{\x}{\tau_{\p,\m} \leq n \log n + n \log(3/\varepsilon)} \leq c/n,
$$
for a suitable constant $c = c(\varepsilon)$.}
%%PINO: a che serve \hittingorStm?
\begin{lemma}\label{lem:hitting_or}
Let $\tau_{\p,\m}$ denote the hitting time of either profile $\p$ or profile $\m$ for
the logit dynamics with parameter $\beta$ of the $n$-player pure coordination game.
Then, for every profile $\x$ with at least one player adopting action $+1$ and
                                  at least one player adopting action $-1$,
$$
\Prob{\x}{\tau_{\p,\m} \leq n \log \frac{n}{\varepsilon}} \leq \frac{c}{n},
%\Prob{\x}{\tau_{\p,\m} \leq n \log n + n \log(3/\varepsilon)} \leq c/n,
%PINO I removed 3
$$
for a suitable constant $c = c(\varepsilon)$.
\end{lemma}
\begin{proof}
We start by proving that
$\Prob{\x}{\tau_{\p,\m} \leq t} \leq \Prob{k}{\rho_{0,n} \leq t}$,
where $\rho_{0,n}$ is the hitting time of state $0$ or state $n$ for
the lazy Ehrenfest urn and
$k$ is the number of $+1$ in $\x$.

We say that $\x,\y\in\Omega = \{+1,-1\}^n$ are equivalent if they have the same number
of $+1$'s and we let $Z=\{Z_t\}$ be the \emph{projection} of
the logit dynamics starting from $\x$ over the quotient space
$\Omega_\# = \{0,1, \dots, n\}$.
The hitting time $\tau_{\p,\m}$ for the logit dynamics starting in $\x$
coincide with the hitting time $\hat{\rho}_{0,n}$ of state $0,n \in \Omega_{\#}$
for the projection $Z$ starting in $k$.
$Z$ has the following transition matrix $P_\#$.
\begin{equation}\label{eq:tranproj}
P_\#(i,i-1) = \frac{i}{2n}; \qquad P_\#(i,i) = \frac{1}{2}; \qquad P_\#(i,i+1) = \frac{n-i}{2n}; \qquad\qquad\mbox{ for } i=2,\ldots,n-2
\end{equation}
and
\begin{align*}
 P_\#(1,0) & = P_\#(n-1,n) = \frac{1}{n(1 + e^{-\beta})} \leq \frac{1}{n};\\
 P_\#(1,1) & = P_\#(n-1,n-1) = \frac{n-1}{2n} + \frac{1}{n(1 + e^{\beta})};\\
 P_\#(1,2) & = P_\#(n-1,n-2) = \frac{n-1}{2n}.
\end{align*}
$P_\#$ differs from the transition matrix of the lazy Ehrenfest urn only for states
$1$ and $n-1$.
Observe though that the transitions from state $1$ to state $0$ and from state $n-1$
to state $n$ in $Z$ have smaller probability than in the Ehrenfest urn and
this can only increase the hitting time  of $0$ and $1$ in $Z$.

The lemma then follows from Lemmas~\ref{lemma:ERub} and~\ref{lemma:lazyslowdown}.
\end{proof}

In the next lemma we show that, if the chain starts from a state containing at least one $1$ and at least one $0$,
then after $\OO(n \log n)$ time steps the distribution of the chain is $\varepsilon$-close to the distribution $U$.

\begin{lemma}\label{lem:polyOR}
Let $R=\{+1,-1\}^n\setminus\{\p,\m\}$ and let $U$ be the uniform distribution over $R$.
Then for the logit dynamics for the $n$-player pure coordination game, it holds that
$$\pmt{U}{R}{\varepsilon} = \OO\left(n \log \frac{n}{\varepsilon}\right).$$
\end{lemma}
\begin{proof}
Let $\{X_t\}$ be the Markov chain starting at $\x \in R$ and let $\{Y_t\}$ be a lazy random walk on the $n$-cube starting at the uniform distribution $U^\star$ on $\{+1,-1\}^n$, so that $X_t$ is distributed according to $P^t(\x, \cdot)$ and $Y_t$ is uniformly distributed over $\{+1,-1\}^n$. Consider the following \emph{coupling} $(X_t,Y_t)$ (for a formal definition and some useful facts about the coupling of Markov chains, we refer the reader to Appendix~\ref{apx::mcsummary}): when chain $\{X_t\}$ is at state $\y \in \{+1,-1\}^n$ then choose a position $i \in [n]$ uniformly at random and, by denoting with $|\y|$ the number of players playing strategy $+1$ in $\y$,
\begin{itemize}
\item If $2 \leq |\y| \leq n-2$, or $|\y| = 1$ and $X_t$ has $-1$ in position $i$, or $|\y| = n-1$ and $X_t$ has $+1$ in position $i$,
then choose an action $a \in \{-1,1\}$ uniformly at random and update both chains $X_t$ and $Y_t$ in position $i$ with action $a$;
\item If $|\y| = 0$ or both $|\y| = 1$ and $X_t$ has $+1$ in position $i$, then
\begin{itemize}
\item update both chains at $-1$ in position $i$ with probability $1/2$;
\item update both chains at $+1$ in position $i$ with probability $1/(1+e^\beta)$;
\item update chain $X_t$ at $-1$ and chain $Y_t$ at $+1$ in position $i$ with probability $1/(1+e^{-\beta}) - 1/2$.
\end{itemize}
\item If $|\y| = n$ or both $|\y| = n-1$ and $X_t$ has $-1$ in position $i$, then
\begin{itemize}
\item update both chains at $+1$ in position $i$ with probability $1/2$;
\item update both chains at $-1$ in position $i$ with probability $1/(1+e^\beta)$;
\item update chain $X_t$ at $+1$ and chain $Y_t$ at $-1$ in position $i$ with probability $1/(1+e^{-\beta}) - 1/2$.
\end{itemize}
\end{itemize}
By construction we have that $(X_t, Y_t)$ is a coupling of $P^t(\x, \cdot)$ and $U^\star$. Then
\begin{equation}
\label{eq:initial_bound}
 \tv{P^t(\x, \cdot) - U} \leq \tv{P^t(\x, \cdot) - U^\star} + \frac{1}{2^{n-1}} \leq \Prob{\x,U^\star}{X_t \neq Y_t} + \frac{1}{2^{n-1}},
\end{equation}
where the last inequality follows from Theorem~\ref{thm:coupling}.
Moreover observe that, if at time $t$ all players have been selected at least once and chain $X_t$ has not yet hit profiles $\p$ or $\m$, then the two random variables $X_t$ and $Y_t$ have the same value. Hence
$$
\Prob{\x,U^\star}{X_t \neq Y_t} \leq \Prob{\x}{\tau_{\p,\m} \leq t  \cup  \eta > t} \leq \Prob{\x}{\tau_{\p,\m} \leq t}  + \Prob{\x}{\eta > t},
$$
where $\tau_{\p,\m}$ is the hitting time of $\p$ or $\m$ for chain $X_t$, and $\eta$ is the first time all players have been selected at least once.

From the coupon collector's argument (see, e.g., \cite[Proposition 2.4]{lpwAMS08}) it follows that for every $t \geq n \log (3n/\varepsilon)$
\begin{equation}\label{eq:ccbound}
\Prob{\x}{\eta > t} \leq \varepsilon/3.
\end{equation}
Hence, for $t = n \log n + n \log(3/\varepsilon)$, by combining~\eqref{eq:initial_bound}, \eqref{eq:ccbound} and Lemma~\ref{lem:hitting_or} it holds that
$$
\tv{P^t(\x, \cdot) - U} \leq \frac{1}{2^{n-1}} + \frac{\varepsilon}{3} + \frac{c}{n} \leq \frac{\varepsilon}{2},
$$
for $n$ sufficiently large.
\end{proof}
Finally, we have the following theorem.
\begin{theorem}
For $\beta = \omega(\log n)$, for every $\varepsilon > 0$ and for every $\x \in \{+1,-1\}^n$,
there exists function $T(n)=T_{\varepsilon,\x}(n)$ and distribution $\mu$ such that
$T(n)$ is super-polynomial in $n$,
$\mu$ is $(\varepsilon,T(n))$-metastable and
pseudo-mixing time $\pmt{\mu}{\{\x\}}{\varepsilon}$ is polynomial in $n$ and in
$\log 1/\varepsilon$.
\end{theorem}
\begin{proof}
For $\x\ne\p,\m$, the theorem follows from Lemma \ref{lem:polyOR}.
Moreover, if the dynamics start from profile $\p$ or $\m$
it immediately reaches the metastable distributions $\pi_{\+}$ or $\pi_{-}$.
\end{proof}
%%PINO

\section{Metastability of Graphical Coordination games}
\label{sec::graphmeta}
Consider $n$ players identified by the vertices of a graph $G = (V,E)$.
For each edge of $G$, we have an instance of the two-strategy coordination game defined above,
played by the endpoints of the edge.
Each player picks a strategy from the set of strategy $\{\pm 1\}$ and uses it
for each two-strategy coordination game in which she is involved.
The utility of a player is the sum of the utilities for each two-strategy coordination game she plays.
Graphical coordination games are often used to model the spread of
a new technology in a social network \citep{youngTR00, msFOCS09} with
strategy $+1$ corresponding to adopting the new technology.
It is easy to see that the graphical coordination game on a graph $G$
is a potential game with potential function
$\Phi(\x) = \sum_{e=(u,v) \in E} \Phi_e(\x)$, where $\Phi_e(\x)$ is the
potential of the two-strategy coordination game associated with edge $e$.

We start by describing two distributions that turn out to be metastable for any graphical coordination game.
\begin{lemma}
\label{lem:graphical_extreme_metastable}
Consider the logit dynamics with parameter $\beta$ for the coordination game on a graph
$G$ with minimum degree $d_{\min}$. Then, for any $\varepsilon > 0$,
the distribution $\pi_{\+}$ concentrated in profile $\p=(+1)^n$
is $(e^{-\beta\cdot\Delta \cdot d_{\min}}, 1)$-metastable.
Similarly, distribution $\pi_{\meno}$ concentrated in profile $\m=(-1)^n$
is $(e^{-\beta\cdot\delta \cdot d_{\min}}, 1)$-metastable.
\end{lemma}
\begin{proof}
The bottleneck ratio of $\p$ is
$$ \Bn(\p)=
	\frac{\sum_{i = 1}^n \pi(\p) P({\p},(\p_{-i}, -1))}{\pi(\p)}
        = \sum_{i=1}^n \frac{1}{n} \frac{1}{1 + e^{-\beta(\Phi(\p) - \Phi(\p_{-i}, -1))}}.$$
An edge adjacent to player $i$ has potential $\Delta$ in $\p$,
whereas in $(\p_{-i},-1)$ it has potential $0$.
Therefore $\Phi(\p)-\Phi(\p_{-i}, -1)=\Delta \cdot d_i$, where $d_i$ is the degree of $i$.
Hence
$$
 \Bn(\p)= \sum_{i=1}^n \frac{1}{n} \frac{1}{1 + e^{\beta \Delta \cdot d_i}} \leq \frac{1}{1 + e^{\beta \cdot\Delta \cdot d_{\min}}}.
$$
Thus by Lemma~\ref{lem:meta:bottleneck}, we have that $\pi_{\+}$ is
$\left((1 + e^{\beta \cdot\Delta \cdot d_{\min}})^{-1},1\right)$-metastable and the theorem follows.
The metastability of $\pi_{\meno}$ can be proved in similar way.
\end{proof}
The above result should not be surprising as it simply says that if players happen to be coordinated
then they will stay so for exponentially long (in the parameter $\beta$).
This however does not conclude the study of metastability of graphical coordination games
as nothing is said about the behavior of the logit dynamics for different initial
states. More specifically, the following question remains unanswered:
do we have for each (class of) initial state a metastable distribution
that is quickly reached?
That is we look for $(\epsilon,T)$-metastable distributions with small $\epsilon$,
large $T$ and small pseudo-mixing time.
We positively answer this question for the two extremal graphs in terms of connectivity:
the complete graph and the ring (that are the
graphical coordination games that have been firstly studied in \cite{ellisonECO93}).

\subsection{The Curie-Weiss game}\label{sec::isingmeta}
The \emph{Curie-Weiss game} is the game-theoretic formulation of the well-studied Curie-Weiss model
(the Ising model on the complete graph),
used by physicists to model the interaction between magnets in
a ferro-magnetic system~\citep{Mart1999}.
This game is also a \emph{population game}.
Hence, the metastable states of the logit dynamics for this game has been object of intense analysis by both physicists and economists
\cite{begkPTRF01,sandholmMIT10,THEC:THEC187}.
Our approach complements these works by focusing on metastable distributions.

Specifically, the Curie-Weiss game is a special graphical coordination game
where the graph $G$ is a clique and each player is involved in a two-strategy coordination game
with $a=b=1$ and $c=d=-1$ with every other player.
Note that for this game the utility of player $i$
at profile $\x = (x_1, \dots, x_n) \in \{\pm 1\}^n$
can be written as $u_i(\x) = x_i \sum_{j \neq i} x_j$.
Similarly, the potential function $H$ can be written as $H(\x) = \sum_{(j,k)} x_j x_k$.
Thus, the Markov chain of the logit dynamics for the Curie-Weiss game with parameter $\beta$
has stationary distribution $\pi(\x) = e^{\beta H(\x)}/Z$,
where $Z=\sum_{\y}e^{\beta H(\y)}$ is the partition function.

The \emph{magnetization} of $\x$ is defined as $S(\x) = \sum_{i=1}^n x_i$ and
we observe that the potential of a profile $\x$ depends only on its magnetization.
More precisely, if $S(\x) = k$ then $H(\x) = H(k):= \frac{1}{2}\left( k^2 - n \right)$.
To see this, let us name $p$ and $m$ the number of $+1$ and $-1$ respectively, in profile $\x$,
and observe that $p - m = S(\x) = k$ and $p + m = n$.
Each pair of players with the same sign contributes for $+1$ to $H(\x)$ and
each pair of players with opposite signs contributes for $-1$;
since there are $\binom{p}{2}$ pairs where both players play $+1$,
$\binom{m}{2}$ pairs where both play $-1$ and
$p\cdot m$ pairs where players play opposite strategies, we have that
$$
H(\x) = \binom{p}{2} + \binom{m}{2} - p \cdot m = \frac{1}{2}((p-m)^2 - (p+m)).
$$

The mixing time of the logit dynamics for the Curie-Weiss game
is known to be exponential for every $\beta > 1/n$ (see Theorem~15.3 in \cite{lpwAMS08}).
In this section we show that, if the Markov chain starts from a profile where the number of $+1$ (respectively $-1$)
is a sufficiently large majority and if $\beta$ is large enough then,
after a polynomial \emph{pseudo-mixing} time, the distribution of the chain is close, in total variation distance, to
$\pi_{\+}$ (respectively, $\pi_{\meno}$).
Note that from Lemma~\ref{lem:graphical_extreme_metastable} and Lemma~\ref{lem:meta:1}, since $\Delta=2$ and $d_{\min} = n-1$, it follows that
$\pi_{\+}$ and $\pi_{\meno}$ are metastable for a time longer than any polynomial for any $\beta = \omega(\log n / n)$.

\medskip
Let $X_t$ be the logit dynamics for the Curie-Weiss game, and consider the magnetization process $S_t := S(X_t)$. Observe that $S_t$ is itself a Markov chain, with state space $\Omega = \{ -n, -n+2, \cdots, n-4, n-2, n \}$. When at state $k \in \Omega$, the probability to go right (to state $k+2$) or left (to state $k-2$) is respectively
\begin{equation}\label{eq:bdrates}
\Prob{k}{S_1 = k+2} = p_k = \frac{n-k}{2n} \frac{1}{1+e^{- 2(k+1) \beta}}; \qquad
\Prob{k}{S_1 = k-2} = q_k = \frac{n+k}{2n} \frac{1}{1 + e^{2 (k-1) \beta}}.
\end{equation}
Indeed, let us evaluate the probability to jump from a profile $\x$ with magnetization $k$ to a profile with magnetization $k+2$. If $S(\x) = k$ then there are $(n+k)/2$ players playing $+1$ and $(n-k)/2$ players playing $-1$. The chain moves to a profile with magnetization $k+2$ if a player playing $-1$ is selected, this happens with probability $(n-k)/2n$, and she updates her strategy to $+1$, this happens with probability
$$
\frac{e^{\beta u_i(\x_{-i},+1)}}{e^{\beta u_i(\x_{-i},+1)} + e^{\beta u_i(\x_{-i},-1)}} = \frac{1}{1 + e^{\beta \left[ u_i(\x_{-i},-1) - u_i(\x_{-i},+1) \right]}}.
$$
Finally observe that $u_i(\x_{-i},-1) - u_i(\x_{-i},+1) = -2 \sum_{j \neq i} x_j = -2 \left( S(\x) - x_i \right) = -2 (k+1)$.

For $a,b \in [-n,n]$, with $a < b$, let $\tau_{a,b}$ be the random variable indicating the first time the magnetization chain reaches a state $k$ with $k \leq a$ or $k \geq b$, that is
$$
\tau_{a,b} = \min \left\{ t \in \mathbb{N} \colon S_t \leq a \mbox{ or } S_t \geq b \right\}.
$$
At time $\tau_{a,b}$, chain $S_{\tau_{a,b}}$ can be in one out of two states, namely the largest state smaller than or equal to $a$ or the smallest state larger than or equal to $b$.
The following technical lemmas give an upper bound on the probability that when the chain exits from interval $(a,b)$, it happens on the left side of the interval.

In the first lemma we show that, if the chain starts from a sufficiently large positive state $k$, and if $\beta k^2 \geq c \log n$ for a suitable constant $c$, then when chain $S_t$ gets out of interval $(0,n/2)$, it happens on the $n/2$ side with high probability.
\newcommand{\maghitbeforehalfStm}{Let $k \in \Omega$ be the starting state with $4 \leq k \leq n/2$. If $\beta \geq 6/n$ and $\beta k^2 \geq 16 \log n$, then
$$\Prob{k}{S_{\tau_{0,n/2}} \leq 0} \leq 1/n.$$}
\begin{lemma}\label{lemma:maghit0beforehalf}
\maghitbeforehalfStm
\end{lemma}
\begin{proof}
According to \eqref{eq:bdrates}, the ratio of $q_h$ and $p_h$ is
$$
\frac{q_h}{p_h} = \frac{n+h}{n-h} \cdot \frac{1+e^{-2(h+1)\beta}}{1 + e^{2(h-1)\beta}}.
$$
Now observe that for all $h \geq 2$ it holds that
\begin{equation}\label{eq:ubsingletranratio}
\frac{1+e^{-2(h+1)\beta}}{1 + e^{2(h-1)\beta}} \leq e^{-2 (h-1) \beta} \leq e^{-h \beta},
\end{equation}
and for all $h \leq n/2$ it holds that
$$
\frac{n+h}{n-h} = \frac{1+h/n}{1-h/n} \leq e^{3h/n}.
$$
Hence, for every $2 \leq h \leq n/2$ we can give the following upper bound
\begin{equation}
\label{eq:q_h_less_p_h_h_small}
\frac{q_h}{p_h} \leq e^{3h/n} \cdot e^{-\beta h} = e^{-(\beta - 3/n)h} \leq e^{-\frac{1}{2} \beta h},
\end{equation}
where in the last inequality we used $\beta \geq 6/n$.

Thus, for each state $h$ of the chain with $k/2 \leq h \leq n/2$ we have that the ratio $q_h/p_h$ is less than $e^{-\frac{1}{4} \beta k}$. If the chain starts at $k$, by applying Lemma~\ref{lemma:hit0beforen} it follows that the probability of reaching $k/2$ before reaching $n/2$ is less than $\left(e^{-\frac{1}{4} \beta k}\right)^\ell$, where $\ell$ is the number of states between $k/2$ and $k$, that is $\ell = k/4$.  Hence, for every $4 \leq k \leq n/2$, if $\beta k^2 \geq 16 \log n$, the chain starting at $k$ hits state $n/2$ before state $k/2$ with probability
$$
\Prob{k}{S_{\tau_{k/2,n/2}} \leq k/2} \leq e^{-\frac{1}{16}\beta k^2} \leq \frac{1}{n}.
$$
The thesis follows by observing that $\Prob{k}{S_{\tau_{0,n/2}} \leq 0} \leq \Prob{k}{S_{\tau_{k/2,n/2}} \leq k/2}$.
\end{proof}

In the next lemma we show that, if the chain starts from a state $k \geq n/2$, and if $\beta \geq c \log n /n$ for a suitable constant $c$, then when chain $S_t$ reaches one of the endpoints of interval $(0,n)$ it is on the $n$ side with probability exponentially close to $1$.
\newcommand{\maghitbeforenStm}{Let $k \in \Omega$ be the starting state with $n/2 \leq k \leq n-1$. If $\beta \geq 8 \log n /n$, then
$$
\Prob{k}{S_{\tau_{0,n}} \leq 0} \leq (2/n)^{n/8}.
$$}
\begin{lemma}\label{lemma:maghit0beforen}
\maghitbeforenStm
\end{lemma}
\begin{proof}
Observe that for every $h \leq n-1$ it holds that $\frac{n+h}{n-h} \leq 2n$, and by using it together with \eqref{eq:ubsingletranratio} we have that $q_h / p_h \leq 2n e^{-\beta h}$ for every $2 \leq h \leq n-1$. Thus, for every $k/2 \leq h \leq n-1$ it holds that
\begin{equation}
 \label{eq:q_h_less_p_h_h_large}
 q_h / p_h \leq 2n e^{-\frac{1}{2} \beta k} \leq 2/n,
\end{equation}
where in the last inequality we used $k \geq n/2$ and $\beta \geq 8 \log n / n$. Hence, if the chain starts at $k$, by applying Lemma~\ref{lemma:hit0beforen} it follows that the probability of reaching $k/2$ before reaching $n$ is less than $(2/n)^\ell$, where $\ell = k/4 \geq n/8$ is the number of states between $k/2$ and $k$. Hence,
\[
\Prob{k}{Y_{\tau_{0,n}} \leq 0} \leq \Prob{k}{Y_{\tau_{k/2,n}} \leq k/2} \leq (2/n)^{n/8}.\qedhere
\]
\end{proof}

Finally, in the next lemma we show that for every starting state between $0$ and $n$, the expected time the chain reaches $0$ or $n$ is at most $\OO(n^3)$.
\newcommand{\expecexitStm}{For every $k \in \Omega$ with $k \geq 0$ it holds that $\Expec{k}{\tau_{0,n}} \leq n^3$.}
\begin{lemma}\label{lemma:expecexit}
\expecexitStm
\end{lemma}
\begin{proof}
Consider the birth-and-death chain $\{Y_t^\star\}$ on $\Omega_{+} = \{l \in \Omega \mid l \geq 0\}$ and let $\tau_n^\star$ the hitting time of $n$ in this chain. It is obvious that $\Expec{k}{\tau_{0,n}} \leq \Expec{k}{\tau^\star_{n}}$. It is well-known (see, for example, Section~2.5 in \citep{lpwAMS08}) that
$$
 \Expec{k}{\tau^\star_{n}} = \sum_{l = \frac{k+2}{2}}^{n/2} \frac{1}{q_{2l} w_{2l}} \sum_{j=\frac{n \mod 2}{2}}^{l-1} w_{2j},
$$
where $w_{n \mod 2} = 1$ and $w_{2j} = \prod_{i=1}^j p_{2(j-1)}/q_{2j}$. From simple computations, we obtain
\begin{align*}
 \Expec{k}{\tau^\star_{n}} & = \sum_{l = \frac{k+2}{2}}^{n/2} \sum_{j=\frac{n \mod 2}{2}}^{l-1} \frac{1}{p_{2j}} \prod_{i=j+1}^{l-1} \frac{q_{2i}}{p_{2i}}\\
& \leq \sum_{l = \frac{k+2}{2}}^{n/2} \sum_{j=\frac{n \mod 2}{2}}^{l-1} \frac{1}{p_{2j}},
\end{align*}
where the inequality follows from \eqref{eq:q_h_less_p_h_h_small} and \eqref{eq:q_h_less_p_h_h_large}. Finally, the Lemma follows by observing that $p_{2j} \geq \frac{1}{2n}$ for every $j \geq 0$.
\end{proof}

Now we can state and prove the main theorem of this section.
\begin{theorem}\label{theorem:metalargebeta}
Let $\x$ be a profile of magnetization $S(\x) = k$. If $\beta \geq 8 \log n / n$ and $k^2 > 16 \log n / \beta$ then
$\pmt{\pi_{\+}}{\{\x\}}{\varepsilon} \leq n^4$ if $k$ is positive and $\pmt{\pi_{\meno}}{\{\x\}}{\varepsilon} \leq n^4$ if $k$ is negative.
\end{theorem}
\begin{proof}
Consider without loss of generality the case of starting state with positive magnetization, $S(\x) = k \geq 0$. Let $\tau_n$ be the first time the chain hits state with all $+1$ and recall that $\tau_{0,n}$ is the first time the magnetization of the chain is either $n$ or less than or equal to $0$.
% $$
% \tau_{0,n} = \min \left\{ t \in \mathbb{N} \colon S_t = n \mbox{ or } S_t \leq 0 \right\}.
% $$
Since $\{ \tau_n > t,  \tau_{0,n} \leq t\}$ implies that the magnetization chain reaches $0$ before reaching $n$ we have
\begin{align*}
\Prob{\x}{\tau_{n} > t} & = \Prob{\x}{\tau_{n} > t,  \tau_{0,n} > t} + \Prob{\x}{\tau_{n} > t,  \tau_{0,n} \leq t} \\
& \leq  \Prob{\x}{\tau_{0,n} > t} + \Prob{\x}{S_{\tau_{0,n}} \leq 0} \\
& \leq \frac{\Expec{\x}{\tau_{0,n}}}{t} + \Prob{\x}{S_{\tau_{0,n}} \leq 0}.
\end{align*}
As for the first term of the sum, from Lemma~\ref{lemma:expecexit} it follows that $\Expec{\x}{\tau_{0,n}}/t \leq 1/n$ for $t \geq n^4$. As for the second term, by conditioning on the position of the chain when it gets out of subinterval $(0,n/2)$ we have
\begin{align*}
\Prob{k}{S_{\tau_{0,n}} \leq 0} & = \Prob{k}{S_{\tau_{0,n}} \leq 0 \mid Y_{\tau_{0,n/2}} \leq 0 }\Prob{k}{S_{\tau_{0,n/2}} \leq 0} + \\
& \quad +\Prob{k}{S_{\tau_{0,n}} \leq 0 \mid S_{\tau_{0,n/2}} \geq n/2 }\Prob{k}{S_{\tau_{0,n/2}} \geq n/2} \\
& \leq \Prob{k}{S_{\tau_{0,n/2}} \leq 0} + \Prob{k}{S_{\tau_{0,n}} \leq 0 \mid S_{\tau_{0,n/2}} \geq n/2}.
\end{align*}
From Lemma~\ref{lemma:maghit0beforehalf} we have that $\Prob{k}{S_{\tau_{0,n/2}} \leq 0} \leq 1/n$, and observe that
$$
 \Prob{k}{S_{\tau_{0,n}} \leq 0 \mid S_{\tau_{0,n/2}} \geq n/2} \leq \Prob{n/2}{S_{\tau_{0,n}} \leq 0} \leq (2/n^{n/8}),
$$
where the last inequality follows from Lemma~\ref{lemma:maghit0beforen}. Hence for every $t \geq n^4$ it holds that  $\Prob{\x}{\tau_{n} > t} \leq 3/n$. The thesis then follows.
\end{proof}

\subsection{Graphical Coordination games on rings}\label{sec::coordmeta}
In this section we study graphical coordination games on a ring.

In \citep{afpppSPAA11} it is showed that the mixing time of the logit dynamics for this game is $\Omega\left(e^{2\delta\beta}\right)$.
If $\Delta>\delta$,
techniques of \citep{bkmpPTRF05} can be generalized to obtain an upper bound to the mixing time that
is polynomial in $n$ and exponential in $\beta$.
If, instead, $\Delta=\delta$, an almost matching upper bound is given in \citep{afpppSPAA11}.
These results show that
the mixing time is polynomial in $n$ for $\beta= \OO(\log n)$ and
greater than any polynomial in $n$, for $\beta = \omega(\log n)$.
In this section we shall show that, for these large values of $\beta$, despite
the large mixing time, from each starting point the logit dynamics quickly
approach a metastable distribution.
Specifically, our result is summarized by the following theorem.
\begin{theorem}
\label{thm:ringdom_pseudo}
Consider the logit dynamics with parameter $\beta$ for an $n$-player coordination game on a ring.
Then, for $n$ sufficiently large, for $\beta=\omega(\log n)$, for every $\epsilon > 0$ and for every $\x \in \{\pm 1\}^n$,
there is a $(\epsilon, T(n))$-metastable distribution $\mu$, with $T$ super-polynomial in $n$,
such that the pseudo-mixing time $\pmt{\mu}{\{\x\}}{\varepsilon}$ is polynomial in $n$ and in $1/\varepsilon$.
\end{theorem}

% The proof of the theorem is split in two parts:
% in the next subsection we consider games with risk dominant strategies ($\Delta > \delta$); Subsection~\ref{subsec:ringnodom} consider games without risk dominant strategies $(\Delta = \delta)$.
%
% \subsubsection{Games with risk dominant strategies}
% \label{subsec:ringdom_pseudo}
% We consider first the case of graphical coordination games with risk dominant strategies (that is, $\Delta > \delta$).
We will show that, depending on the starting state $\x$,
the Markov chain $\mathcal{M}_\beta$ of the logit dynamics with parameter $\beta$ for the $n$-player coordination game on the ring,
rapidly approaches a $(\epsilon,\epsilon \cdot e^{2\beta\delta})$-metastable distribution.
Specifically, our results confirm the findings of \cite{ellisonECO93,Ellison2000},
that the logit dynamics for this game tend to reach in the long run the metastable states,
namely, the profiles in which either all players adopt strategy $-1$ or they all adopt with strategy $+1$.
Still our approach based on metastable distributions allows us to have a finer description of the system in the medium run.
For example, our results for graphical coordination games on the ring without a risk-dominant strategy allow us to understand that
the probability that one of these states is reached essentially depends only on the number of players that adopt a given strategy at beginning,
and not on their location on the ring.

The proof of the theorem separately considers games with risk dominant strategies (i.e., $\Delta > \delta$) and games without risk dominant strategies (i.e., $\Delta = \delta$).
As for the first ones,
let us first introduce some useful notation.
We define $R \subseteq \{\pm 1\}^n$ as the set of profiles in which at least
two {\em adjacent} players are playing $+1$.
Moreover we define $S_d \subseteq \{\pm 1\}^n$ as the set of profiles
where \emph{exactly} $d$ non-adjacent players are playing $+1$,
$S_{\geq d} = \bigcup_{i=d}^n S_i$
and $S_d^\star = S_{\geq d} \cup R$.
We remind that $\p$ denotes the profile where all players are playing $+1$
and $\m$ denotes the profile where all players are playing $-1$.

Our proof is based on the following three technical lemmas whose proofs are postponed to Appendix~\ref{app:pfromR}, Appendix~\ref{app:pfromSd}
and Appendix~\ref{apx:pmfromSd}, respectively.
The first two lemmas provide an upper bound to the probability that $\mathcal{M}_\beta$
takes too long to hit profile $\p$ when starting from $\x\in R$ or from
$\x\in S_d$. The last lemma, instead, provides an upper bound to the probability that $\mathcal{M}_\beta$
takes too long to hit one of the two profiles $\p$ and $\m$ for any starting state.

More precisely, recalling that $\tau_\p$ denotes the hitting time of profile $\p$,
and $\tau_{\{\p, \m\}}$ denotes the hitting time of the set of profiles $\{\p, \m\}$,
we have that

\newcommand{\pfromRStm}{For all $\x\in R$ and $\epsilon>0$, if $\Delta > \delta$, $\beta=\omega(\log n)$ and $n$ is sufficiently large
$$P_{\x}\left(\tau_\p>\frac{8-\epsilon}{\epsilon}\cdot n^2\right)\leq \frac{\epsilon}{4}.$$}
\begin{lemma}
\label{lem:pfromR}
\pfromRStm
\end{lemma}

\newcommand{\pfromSdStm}{For all $d>0$, $\x\in S_d$ and $\epsilon>0$, if $\Delta > \delta$, $\beta=\omega(\log n)$ and $n$ is sufficiently large
$$P_{\x}\left(\tau_\p>\frac{8}{\epsilon}\cdot n^2\right)\leq \frac{1}{2^d + 1}+\frac{\epsilon}{4}.$$}
\begin{lemma}
\label{lem:pfromSd}
\pfromSdStm
\end{lemma}

\newcommand{\pmfromSdStm}{For all $\x\in \{\pm 1\}^n$ and $\epsilon>0$, if $\Delta > \delta$, $\beta=\omega(\log n)$ and $n$ is sufficiently large
$$P_{\x}\left(\tau_{\{\p,\m\}}>\frac{8}{\epsilon}\cdot n^2\right) \leq \frac{\varepsilon}{2}.$$}
\begin{lemma}
\label{lem:pmfromSd}
\pmfromSdStm
\end{lemma}

As for the case that no risk dominant strategy exists, the proof is based on a technical lemma, whose proof is postponed to Appendix~\ref{app:proofsw}, that shows that with high probability the hitting time of the set $\{\p, \m \}$ is polynomial in the number of players.
\newcommand{\rndtaupmStm}{For all $\x\in \{\pm 1\}^n$ and $\epsilon>0$, if $\Delta = \delta$, $\beta=\omega(\log n)$ and $n$ is sufficiently large
$$\Prob{\x}{\tau_{\{\p,\m\}} > n^5} \leq o(1).$$}
\begin{lemma}
\label{lem:ringnodom_taupm}
\rndtaupmStm
\end{lemma}

We are now ready to prove Theorem \ref{thm:ringdom_pseudo}.
\begin{proof}[Proof (of Theorem \ref{thm:ringdom_pseudo}).]
We start showing that for each possible initial state, the pseudo-mixing time of the Markov chain $\mathcal{M}_\beta$ to a $(\varepsilon, \varepsilon \cdot T(n))$-metastable distribution, for a function $T$ super-polynomial in $n$, is polynomial in $n$ and in $1/\varepsilon$. In particular, we will show that $\mathcal{M}_\beta$ approaches a convex combination of $\pi_{\+}$ and $\pi_{\meno}$.
These two distributions, from Lemma~\ref{lem:graphical_extreme_metastable} and since in a ring any vertex has degree 2,
are $(e^{-2\delta \beta}, 1)$-metastable.
Then, from Lemma~\ref{lem:meta:1} it follows that they are $(\varepsilon, \varepsilon \cdot e^{2\delta \beta})$-metastable.
Consequently, by Lemma~\ref{lem:comb}, any their convex combination is $(\varepsilon, \varepsilon \cdot e^{2\delta \beta})$-metastable.
Finally, since $\beta=\omega(\log n)$, it follows that $\varepsilon \cdot e^{2\delta \beta}$ is super-polynomial in $n$ as desired.

For the pseudo-mixing time we distinguish three cases.
\begin{description}
 \item[$\Delta > \delta$, $\x \in S^\star_{d}$ and $d \geq \log_2\left(\frac{4}{\epsilon} - 1\right)$:]
 We prove that the pseudo-mixing time $\pmt{\pi_{\+}}{S^\star_{d}}{\epsilon}$
 to $\pi_{\+}$ from any $\x \in S^\star_{d}$, with $d \geq \log_2\left(\frac{4}{\epsilon} - 1\right)$,
 is polynomial in $n$ and $1/\varepsilon$.
 Indeed, by Lemma~\ref{lem:pfromR} and Lemma~\ref{lem:pfromSd} we have that
$$\max_{x\in S^\star_{d}}
P_\x\left(\tau_\p>\frac{8}{\epsilon}\cdot n^2\right)\leq \frac{\epsilon}{2}.$$
Moreover, from Lemma~\ref{lem:graphical_extreme_metastable} and Lemma~\ref{lem:meta:1}
it follows that $\pi_{\+}$ is $(\varepsilon/2, \varepsilon/2 \cdot e^{2\delta \beta})$-metastable.
Thus, for sufficiently large $n$, since $\frac{8}{\epsilon}\cdot n^2\leq \frac{\epsilon}{2}\cdot e^{2\delta \beta}$,
$\pi_{\+}$ is $\left(\frac{\varepsilon}{2}, \frac{8}{\epsilon}\cdot n^2\right)$-metastable.
Then, by Lemma~\ref{lem:pmt:hitting}, we obtain that
$$d_{\pi_{\+}}^{S^\star_{d}}\left(\frac{8}{\epsilon}\cdot n^2\right)\leq \epsilon.$$
Then, from the definition of pseudo-mixing time, we obtain that
$$\pmt{\pi_{\+}}{S^\star_{d}}{\epsilon}\leq \frac{8}{\epsilon}\cdot n^2.$$

 \item[$\Delta > \delta$, $\x \in S_d$ and $d < \log_2\left(\frac{4}{\epsilon} - 1\right)$:]
In this case we prove that the pseudo-mixing time to the distribution
$$\mu_\x = \alpha_\x \pi_{\+} + (1 - \alpha_\x) \pi_{\meno},$$
where
\begin{equation}
\label{eq:alpha}
 \alpha_\x = \Prob{\x}{\tau_\p \leq \tau_\m \mid \tau_{\p,\m} \leq \frac{8}{\epsilon} \cdot n^2}
\end{equation}
is polynomial.
Indeed from Lemma \ref{lem:pmfromSd} we have that
$$
 P_\x\left(\tau_{\{\p,\m\}} > \frac{8}{\epsilon} \cdot n^2\right) \leq \frac{\epsilon}{2}.
$$
Moreover, from Lemma~\ref{lem:graphical_extreme_metastable} and Lemma~\ref{lem:meta:1}
it follows that $\pi_{\+}$ and $\pi_{\meno}$ are both $(\varepsilon/2, \varepsilon/2 \cdot e^{2\delta \beta})$-metastable.
Thus, for sufficiently large $n$, since $\frac{8}{\epsilon}\cdot n^2\leq \frac{\epsilon}{2}\cdot e^{2\delta \beta}$,
$\pi_{\+}$ and $\pi_{\meno}$ are both $\left(\frac{\varepsilon}{2}, \frac{8}{\epsilon}\cdot n^2\right)$-metastable.
Then, from Lemma~\ref{lem:hitting_convex} we obtain that
$$d_{\mu_{\x}}^{\{\x\}}\left(\frac{8}{\epsilon}\cdot n^2\right)\leq \epsilon.$$
Finally, from the definition of pseudo-mixing time, we have that
\[\pmt{\mu_\x}{\{\x\}}{\epsilon}\leq \frac{8}{\epsilon}\cdot n^2.\]

\item[$\Delta = \delta$:]
This case can be proved similarly to the previous case, by using Lemma~\ref{lem:ringnodom_taupm} in place of Lemma \ref{lem:pmfromSd}.
It is surprising that, if $\Delta = \delta$,
the coefficient $\alpha_\x$ defined in \eqref{eq:alpha},
for any $\x \in \{\pm 1\}^n$,
depends ``almost'' only on the number of players
selecting strategy $+1$ in $\x$
(see Lemma~\ref{lem:ringnodom_conv} in Appendix~\ref{apx:alpha}).
Thus, for example, the metastable behavior
of profiles with $d$ adjacent players with strategy $+1$
is almost the same as profiles where the same players are far from each other.
\qedhere
\end{description}
\end{proof}

\section{Conclusions and open problems}\label{sec::conclusions}
Logit dynamics are clean and tractable dynamics that well model the behaviour of limited-rationality players in a strategic game. The stationary distribution of the induced Markov chain is the natural long-term equilibrium concept for games under logit dynamics. However, when the mixing time is long, the behavior of the Markov chain in the transient phase becomes important and it is worth looking for ``regularities'' at a time-scale shorter than mixing time. Such regularities have been previously explored, for some classes of Markov chains, by means of ``metastable states''. We believe that a more general and useful concept is that of ``metastable distributions''.

In this paper we defined a quantitative notion of metastable distribution and we analyzed the metastability properties of the logit dynamics for some classes of coordination games. We showed that, even when the mixing time is exponential, it is often possible to find some distributions that well-approximate the distribution of the chain for a time-window of super-polynomial size. Such metastable distributions can be found even in the case of the OR-game, where no partition of the state space in metastable \emph{states} exists. A natural open question is whether the metastability properties for coordination games we observed in this paper hold in general for potential games.

In the case of the Curie-Weiss model on the complete graph, we showed that when $\beta > c \log n /n$ the two degenerate distributions are metastable for $\poly(n)$ time and they are quickly reached from a large fraction of the state space.
We note that the metastability properties when $1/n < \beta < \log n /n$ have been investigated in \cite{ferraioli2015metastability},
where it is proved that, even in this case, for every starting profile the logit dynamics quickly converge to a distribution that is
metastable for long time.
% It would be interesting to investigate the metastability properties when $1/n < \beta < \log n /n$. Indeed, in that range the mixing time is exponential but the distributions concentrated in the two extremal states are not metastable.

\smallskip
\paragraph{Acknowledgment.} We wish to thank Paolo Penna for useful ideas, hints, and discussions.
We also thank the anonymous referees for their comments and suggestions.

\bibliographystyle{plainnat}
\bibliography{logit}

\newpage
\appendix
\begin{center}
\begin{LARGE}
\textbf{Appendix}\\
\end{LARGE}
\end{center}

\section{Markov chain summary}
\label{apx::mcsummary}
In this section we recall some basic facts about Markov chains.
For a more detailed treatment,
we refer the reader to~\citep{lpwAMS08}.

\paragraph{Total variation distance.}
The {\em total variation} distance
$\tv{\mu - \nu}$
between two probability distributions
$\mu$ and $\nu$ on $\Omega$ is defined as
$$
\tv{\mu-\nu}:=\max_{A\subset\Omega} |\mu(A)-\nu(A)| = \frac{1}{2} \sum_{x\in\Omega} |\mu(x)-\nu(x)|.$$

The total variation distance is actually a distance and, in particular,
the triangle inequality holds. That is, the following simple fact holds.
\begin{fact}
\label{fact:triangleTV}
For distributions $\mu_1,\mu_2,$ and $\mu_3$,
it holds that
$$\tv{\mu_1-\mu_3}\leq\tv{\mu_1-\mu_2}+\tv{\mu_2-\mu_3}.$$
\end{fact}

% \begin{fact}
% \label{fact:alternativeTV}
% The total variation distance of distributions $\mu,\nu$ on $\Omega$ can be written as
% $$\tv{\mu-\nu}=\frac{1}{2} \sum_{x\in\Omega} |\mu(x)-\nu(x)|.$$
% \end{fact}
A stochastic matrix $P$ over $\Omega$ is a non-negative matrix in which rows and columns
are indexed
by elements of $\Omega$ and such that, for all $x\in\Omega$,
it holds that
$$\sum_{y\in\Omega} P(x,y)=1.$$
\begin{fact}
\label{fact:stochasticTV}
For all distributions $\mu$ and $\nu$ on $\Omega$ and all stochastic matrices $P$ it holds that
$$\tv{\mu P-\nu P}\leq\tv{\mu-\nu}.$$
\end{fact}

\paragraph{Mixing time.}
Consider a Markov chain $\mathcal{M}=\{X_t\}$ with {\em finite} state space $\Omega$ and
transition matrix $P$. We stress that $P$ is stochastic matrix over $\Omega$ and we will often identify
$\mathcal{M}$ with $P$.
It is a classical result that if $\mathcal{M}$ is \emph{irreducible} and \emph{aperiodic}\footnote{Roughly speaking, a finite-state Markov chain is irreducible and aperiodic if there exists $t$ such that, for all pairs of states $x,y$, the probability to be in $y$ after $t$ steps, starting from $x$, is positive.} (also called \emph{ergodic})
there exists an unique {\em stationary distribution};
that is, a distribution $\pi$ on $\Omega$ such that $\pi\cdot P=\pi$.

An ergodic Markov chain $\mathcal{M}$
\emph{converges} to its stationary distribution $\pi$;
specifically, there exist constants $C$ and $0<\alpha<1$ such that
$$d(t)\leq C\cdot\alpha^t,$$
where
$$d(t)=
\max_{x\in\Omega}
\tv{P^t(x,\cdot) - \pi}$$
and
$P^t(x,\cdot)$ is the distribution at time $t$
of the Markov chain starting at $x$.
For $0 < \varepsilon \leq 1$,
the \emph{mixing time} is defined as
$$
t_{\text{mix}}(\varepsilon) =
\min \{t\in\mathbb{N}\colon d(t)\leq\varepsilon\}.
$$
It is usual to set $\varepsilon = 1/4$ and to write $\tm$ for $\tm(1/4)$.
\begin{fact}
\label{fact:tm}
For $0<\varepsilon\leq 1$,
$$\tm(\varepsilon)\leq \lceil \log_2\varepsilon^{-1}\rceil\cdot\tm.$$
\end{fact}

\paragraph{Coupling.}
A {\em coupling} of two probability distributions $\mu$ and $\nu$  on
$\Omega$ is a pair of random variables $(X,Y)$ defined on
$\Omega\times\Omega$ such that the marginal distribution of
$X$ is $\mu$ and the marginal distribution of $Y$ is $\nu$.
A well-known property of coupling is given by the following theorem (see, e.g., \cite[Proposition 4.7]{lpwAMS08}).
\begin{theorem}
\label{thm:coupling}
 Let $\mu$ and $\nu$ two probability distributions on $\Omega$. Then
 $$
  \tv{\mu - \nu} = \inf \left\{\Prob{}{X \neq Y} \colon \left(X, Y\right) \text{ is a coupling of $\mu$ and $\nu$}\right\}
 $$
\end{theorem}

%
% A {\em coupling of a Markov chain} $\mathcal{M}$
% with transition matrix $P$ is a process
% $(X_t,Y_t)_{t=0}^\infty$ with the property that
% both $X_t$ and $Y_t$ are Markov chains with transition matrix $P$. When the two coupled chains start at $(X_0,Y_0) = (x,y)$, we write $\Prob{x,y}{\cdot}$ and $\Expec{x,y}{\cdot}$ for the probability and the expectation on the space where the two chains are both defined.
%
% We denote  by $\tc$ the first time the two chains meet; that is,
% $$
% \tc=\min\{t: X_t=Y_t\}.
% $$
% We will consider only couplings of Markov chains with the property that
% for $s\geq\tc$, it holds $X_s=Y_s$.
% The following theorem establish the importance of this tool
% (see, for example, Theorem~5.2 in \citep{lpwAMS08}).
% \begin{theorem}[Coupling]
% \label{thm:coupling}
% Let $\mathcal{M}$ be a Markov chain with finite state space $\Omega$ and
% transition matrix $P$. For each pair of states $x,y\in\Omega$ consider a coupling $(X_t,Y_t)$ of $\mathcal{M}$ with starting states $X_0=x$ and $Y_0=y$.
% Then
% $$
% \tv{P^t(x,\cdot) - P^t(y,\cdot)}\leq \Prob{x,y}{\tc>t}.
% $$
% \end{theorem}

\section{Biased birth-and-death chains}
In this section we consider birth-and-death chains with state space $\Omega = \{0,1, \dots, n\}$ (see Chapter 2.5 in~\citep{lpwAMS08} for a detailed description of such chains). For $k \in \{1, \dots, n-1 \}$ let $p_k = \Prob{k}{X_1 = k+1}$, $q_k = \Prob{k}{X_1 = k-1}$, and $r_k = 1 - p_k - q_k = \Prob{k}{X_1 = k}$. We will be interested in the probability that the chain starting at some state $h \in \Omega$ hits state $n$ before state $0$, namely $\Prob{k}{X_{\tau_{0,n}} = n}$ where $\tau_{0,n} = \min\{ t \in \mathbb{N} \colon X_t \in \{0,n\} \}$.

We start by giving an exact formula for such probability for the case when $p_k$ and $q_k$ do not depend on $k$.
\begin{lemma}\label{lemma:allthesame}
Suppose for all $k \in \{ 1, \dots, n-1 \}$ it holds that $p_k = \epsilon$ and $q_k = \delta$, for some $\varepsilon$ and $\delta$ with $\epsilon + \delta \leq 1$. Then the probability the chain hits state $n$ before state $0$ starting from state $h \in \Omega$ is
$$
\Prob{h}{X_{\tau_{0,n}} = n} =  \frac{1 - \left( \delta / \varepsilon \right)^h}{1 - \left( \delta / \varepsilon \right)^n}.
$$
\end{lemma}
The proof of this result, that is showed below for sake of completeness, uses standard arguments (see, e.g., \cite[p. 159]{taylor2014introduction}).
\begin{proof}
Let $\alpha_k$ be the probability to reach state $n$ before state $0$ starting from state $k$, i.e.
$$
\alpha_k = \Prob{k}{X_{\tau_{0,n}} = n}.
$$
Observe that for $k = 1, \dots, n-1$ we have
\begin{equation}\label{eq:recurpk}
\alpha_k = \delta \cdot \alpha_{k-1} + \varepsilon \cdot \alpha_{k+1} + \left( 1 - (\delta + \varepsilon) \right) \alpha_k.
\end{equation}
Hence
$$
\varepsilon \cdot \alpha_k - \delta \cdot \alpha_{k-1} = \varepsilon \cdot \alpha_{k+1} - \delta \cdot \alpha_k
$$
with boundary conditions $\alpha_0 = 0$ and $\alpha_n = 1$. If we name $\Delta_k = \varepsilon \cdot \alpha_k - \delta \cdot \alpha_{k-1}$ we have $\Delta_k = \Delta_{k+1}$ for all $k$. By simple calculation and using that $\alpha_0 = 0$ it follows that
$$
\alpha_k = \frac{\Delta}{\varepsilon} \sum_{i = 0}^{k-1}\left( \frac{\delta}{\varepsilon} \right)^i = \frac{\Delta}{\varepsilon - \delta} \left( 1 - (\delta / \varepsilon)^k \right).
$$
From $\alpha_n = 1$ we get
$$
\Delta = \frac{\varepsilon - \delta}{\left( 1 - (\delta / \varepsilon)^n \right)}.
$$
Hence
\[
\alpha_k = \frac{1 - \left( \delta / \varepsilon \right)^k}{1 - \left( \delta / \varepsilon \right)^n}.\qedhere
\]
\end{proof}

\begin{lemma}\label{lemma:boundedtransitionrates}
Suppose for all $k \in \{ 1, \dots, n-1 \}$ it holds that $p_k \geq \varepsilon$ and $q_k \leq \delta$, for some $\varepsilon$ and $\delta$ with $\epsilon + \delta \leq 1$. Then the probability to hit state $n$ before state $0$ starting from state $h \in \Omega$ is
$$
\Prob{h}{X_{\tau_{0,n}} = n} \geq  \frac{1 - \left( \delta / \varepsilon \right)^h}{1 - \left( \delta / \varepsilon \right)^n}.
$$
\end{lemma}
\begin{proof}
Let $\{Y_t\}$ be a birth-and-death chain with the same state space as $\{X_t\}$ but transition rates
$$
\Prob{k}{Y_1 = k-1} = \delta; \qquad \Prob{k}{Y_1 = k+1} = \varepsilon.
$$
Consider the following coupling of $X_t$ and $Y_t$: When $(X_t,Y_t)$ is at state $(k,h)$, consider the two $[0,1]$ intervals, each one partitioned in three subintervals as in Fig.~\ref{fig:firstcoupling}.
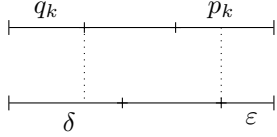
\begin{figure}[ht]
\begin{center}
 \begin{tikzpicture}
  \draw[|-|] (0,1) -- (0.5,1) node[above] {$q_k$} -- (2.8,1) node[above] {$p_k$} -- (3.5,1);
  \draw[|-|] (0,0) -- (0.8,0) node[below] {$\delta$} -- (3.2,0) node[below] {$\varepsilon$} -- (3.5,0);
  \draw[-, dotted] (1,1) -- (1,0);
  \draw[-, dotted] (2.8,1) -- (2.8,0);
  \draw plot[only marks, mark=+] coordinates{(1,1) (2.2,1) (1.5,0) (2.8, 0)};
 \end{tikzpicture}
\caption{Partition for the coupling in Lemma~\ref{lemma:boundedtransitionrates}.}
\label{fig:firstcoupling}
\end{center}
\end{figure}
Let $U$ be a uniform random variable over the interval $[0,1]$ and choose the update for the two chains according to position of $U$ in the two intervals.
Observe that, since $p_k \geq \varepsilon$ and $q_k \leq \delta$, if the two chains start at the same state $h \in \Omega$, i.e.  $(X_0, Y_0) =(h,h)$, then at every time $t$ it holds that $X_t \geq Y_t$. Hence if chain $Y_t$ hits state $n$ before state $0$, then chain $X_t$ hits state $n$ before state $0$ as well. More formally, let $\tau_{0,n}$ and $\hat{\tau}_{0,n}$ be the random variables indicating the first time chains $X_t$ and $Y_t$ respectively hit state $0$ or $n$; then
$$
\left\{Y_{\hat{\tau}_{0,n}} = n \right\} \Rightarrow \left\{X_{\tau_{0,n}} = n \right\}.
$$
Thus
$$
\Prob{h}{X_{\tau_{0,n}} = n } \geq \Prob{h}{Y_{\hat{\tau}_{0,n}} = n } \geq \frac{1 - \left( \delta / \varepsilon \right)^h}{1 - \left( \delta / \varepsilon \right)^n},
$$
where in the last inequality we used Lemma~\ref{lemma:allthesame}.
\end{proof}

\begin{lemma}\label{lemma:hit0beforen}
Suppose for all $k \in \{ 1, \dots, n-1 \}$ it holds that $q_k / p_k \leq \alpha$, for some $\alpha < 1$. Then the probability to hit state $0$ before state $n$ starting from state $h \in \Omega$ is
$$
\Prob{h}{X_{\tau_{0,n}} = 0} \leq \alpha^h.
$$
\end{lemma}
\begin{proof}
Let $\hat{p}_k = \frac{p_k}{p_k + q_k}$ and $\hat{q}_k = \frac{q_k}{p_k + q_k}$ and let $\{Y_t\}$ be the birth-and-death chain with transition rates $\hat{p}_k$ and $\hat{q}_k$.

Let $\{U_t\}$ be an array of random variables such that $U_t = - 1$ with probability $q_{\mbox{\tiny $Y_t$}}$, $U_t = +1$ with probability $p_{\mbox{\tiny $Y_t$}}$ and $U_t = 0$ with remaining probability. We will use $U_t$ to update chains $X_t$ and $Y_t$ at different time steps. Specifically, we denote with $u$ the index of the first variables $U_t$ not used for updating $X_t$ (thus, at the beginning $u = 1$) and : set $Y_{t + 1} = Y_t + U_t$; for chain $X_t$, we toss a coin that gives head with probability $p_{\mbox{\tiny $X_t$}} + q_{\mbox{\tiny $X_t$}}$ and if it gives tail we set $X_{t+1} = X_t$, otherwise we set $X_{t+1} = X_t + U_u$. Roughly speaking, we have that the chain $X_t$ \emph{follows the path} traced by chain $Y_t$: indeed, it is easy to see that, if they start at the same place, the sequence of states visited by the two chains is the same and in the same order. Hence chain $X_t$ hits state $0$ before state $n$ if and only if chain $Y_t$ hits state $0$ before state $n$ and thus $\Prob{k}{X_{\tau^X_{0,n}} = 0}= \Prob{k}{Y_{\tau^Y_{0,n}} = 0}$.

Finally, observe that $\frac{\hat{q}_k}{\hat{p}_k} = \frac{q_k}{p_k} \leq \alpha$ and $\hat{p}_k + \hat{q}_k = 1$. Hence, for every $k \in \{1, \ldots, n-1\}$, we have that $\hat{p}_k \geq \frac{1}{1+\alpha}$ and $\hat{q}_k \leq \frac{\alpha}{1+\alpha}$. This implies, from Lemma~\ref{lemma:boundedtransitionrates}, that, from any state $h \in \Omega$,
$$
 \Prob{h}{Y_{\tau^Y_{0,n}} = n} \geq \frac{1 - \alpha^h}{1 - \alpha^n} \geq 1 - \alpha^h.
$$
The lemma follows.
\end{proof}

\section{Ehrenfest urns}
\label{app:ehrenfest}
The \emph{Ehrenfest urn} is the Markov chain with state space $\Omega = \{ 0,1,\dots, n\}$ that, when at state $k$, moves to state $k-1$ or $k+1$ with probability $k/n$ and $(n-k)/n$ respectively (see, for example, Section~2.3 in~\citep{lpwAMS08} for a detailed description). The next lemma gives an upper bound on the probability that the Ehrenfest urn starting at state $k$ hits state $0$ or $n$ within time step $t$.

\begin{lemma}\label{lemma:ERub}
Let $\{Z_t \}$ be the Ehrenfest urn over $\{0,1, \dots, n\}$ and let $\tau_{0,n}$ be the first time the chain hits state $0$ or state $n$. Then for every $k \geq 1$ it holds that
$$
\Prob{k}{\tau_{0,n} < n \log n + cn} \leq \frac{c'}{n}.
$$
for suitable positive constants $c$ and $c'$.
\end{lemma}
\begin{proof} First observe that for any $t \geq 3$ the probability of hitting $0$ or $n$ before time $t$ for the chain starting at $1$ is only $\OO(1/n)$ larger than  for the chain starting at $2$, which in turn is only $\OO(1/n)$ larger than  for the chain starting at $3$. Indeed, by conditioning on the first step of the chain, we have 
\begin{align*}
\Prob{1}{\tau_{0,n} < t} & = \Prob{1}{\tau_{0,n} < t \mid Z_1 = 0}\Prob{1}{Z_1 = 0} + \Prob{1}{\tau_{0,n} < t \mid Z_1 = 2}\Prob{1}{Z_1 = 2} \\
& = \frac{1}{n} + \frac{n-1}{n} \Prob{2}{\tau_{0,n} < t-1}
\leq \frac{1}{n} + \Prob{2}{\tau_{0,n} < t}
\end{align*}
and
\begin{align*}
\Prob{2}{\tau_{0,n} < t} & = \Prob{2}{\tau_{0,n} < t \mid Z_1 = 1}\Prob{2}{Z_1 = 1} + \Prob{2}{\tau_{0,n} < t \mid Z_1 = 3}\Prob{2}{Z_1 = 3} \\
& = \frac{2}{n} \Prob{1}{\tau_{0,n} < t-1} + \frac{n-2}{n} \Prob{3}{\tau_{0,n} < t-1} \\
& \leq \frac{2}{n} \left(\frac{1}{n} + \Prob{2}{\tau_{0,n} < t} \right) + \frac{n-2}{n} \Prob{3}{\tau_{0,n} < t}.
\end{align*}
Hence,
\begin{align*}
\Prob{2}{\tau_{0,n} < t} & \leq \frac{2}{n-2} + \Prob{3}{\tau_{0,n} < t} \leq \frac{3}{n} + \Prob{3}{\tau_{0,n} < t}; \\
\Prob{1}{\tau_{0,n} < t} & \leq \frac{4}{n} + \Prob{3}{\tau_{0,n} < t}.
\end{align*}
Moreover observe that the probability that the chain starting at $k$ hits state $0$ or $n$ before time $t$ is symmetric,
i.e. $\Prob{k}{\tau_{0,n} < t} = \Prob{n-k}{\tau_{0,n} < t}$,
and it is decreasing for $k$ that goes from $0$ to $n/2$.
In particular, for every $k$ such that $3 \leq k \leq n-3$ it holds that $\Prob{k}{\tau_{0,n} < t} \leq \Prob{3}{\tau_{0,n} < t}$. Now we show that $\Prob{3}{\tau_{0,n} < n \log n + cn} = \OO(1/n) $ and this will complete the proof.

First observe that
$$
 \Prob{3}{\tau_{0,n} < t} = \Prob{3}{\tau_{0,n} < t \wedge \tau_0 < \tau_n} + \Prob{3}{\tau_{0,n} < t \wedge \tau_n < \tau_0},
$$
where $\tau_0$ and $\tau_n$ are the first time the chain hits state $0$ and state $n$, respectively. Then
$$
 \Prob{3}{\tau_{0,n} < t} \leq \Prob{3}{\tau_0 < t} + \Prob{3}{\tau_n < t} \leq \Prob{3}{\tau_0 < t} + \Prob{n-3}{\tau_n < t} = 2 \cdot \Prob{3}{\tau_0 < t},
$$
where the equality follows from the symmetry of the Ehrenfest urn Markov chain.

Let us now consider a path $\mathcal{P}$ of length $t$ starting at state $3$ and ending at state $0$. Observe that any such path must contain the sub-path going from state $3$ to state $0$ whose probability is $6/n^3$. Moreover, for all the other $t-3$ moves we have that if the chain crosses an edge $(i,i+1)$ from left to right then it must cross the same edge from right to left (and vice-versa). The probability for any such pair of moves is
$$
 \frac{n-i}{n} \cdot \frac{i+1}{n} \leq \frac{e^{2/n}}{4},
$$
for every $i$. Hence, for any path $\mathcal{P}$ of length $t$ going from $3$ to $0$, the probability that the chain follows exactly path $\mathcal{P}$ is\footnote{Notice that such probability is zero if $t-3$ is odd.}
$$
\Prob{3}{(X_1, \dots, X_t) = \mathcal{P}} \leq \frac{6}{n^3}\cdot \left( \frac{e^{2/n}}{4}\right)^{(t-3)/2} = \frac{6}{n^3} \cdot \frac{2^3}{e^{3/n}} \cdot \left(\frac{e^{1/n}}{2}\right)^t \leq \frac{48}{n^3} \cdot \left(\frac{e^{1/n}}{2}\right)^t.
$$
Let $\ell$ and $r$ be the number of left and right moves respectively in path $\mathcal{P}$ then $\ell + r = t$ and $\ell - r = 3$. Hence the total number of paths of length $t$ going from $3$ to $0$ is less than
$$
\binom{t}{\ell} = \binom{t}{\frac{t - 3}{2}} \leq 2^t.
$$
Thus, the probability that starting from $3$ the chain hits $0$ for the first time exactly at time $t$ is
$$
\Prob{3}{\tau_0 = t} \leq \binom{t}{\frac{t-3}{2}}\frac{48}{n^3} \cdot \left(\frac{e^{1/n}}{2}\right)^t \leq \frac{48}{n^3} e^{t/n}.
$$
Finally, the probability that the hitting time of $0$ is less than $t$ is
\begin{align*}
\Prob{3}{\tau_0 < t} & \leq \sum_{i = 3}^{t-1} \Prob{3}{\tau_0 = i} \\
& \leq \frac{48}{n^3} \sum_{i = 3}^{t-1} e^{i/n} =  \frac{48}{n^3} \cdot \frac{e^{t/n} - 1}{e^{1/n} - 1} \leq \frac{48 e^c}{n}.
\end{align*}
In the last inequality we used that $e^{1/n} - 1 \geq 1/n$ and $t = n \log n + c n$.
\end{proof}
In the proof of Lemma~\ref{lem:hitting_or} we deal with the lazy version of the Ehrenfest urn. The next lemma, which is folklore, allows us to use the bound we achieved in Lemma~\ref{lemma:ERub} for the non-lazy chain.

\begin{lemma}\label{lemma:lazyslowdown}
Let $\{X_t\}$ be an irreducible Markov chain with finite state space $\Omega$ and transition matrix $P$ and let $\{\hat{X}_t\}$ be its lazy version, i.e. the Markov chain with the same state space and transition matrix $\hat{P} = \frac{P + I}{2}$ where $I$ is the $\Omega \times \Omega$ identity matrix. Let $\tau_A$ and $\hat{\tau}_A$ be the hitting time of the subset of states $A \subseteq \Omega$ in chains $\{X_t\}$ and $\{\hat{X}_t\}$ respectively. Then, for every starting state $b \in \Omega$ and for every time $t \in \mathbb{N}$ it holds that
$$
\Prob{b}{\hat{\tau}_A \leq t} \leq \Prob{b}{\tau_A \leq t}.
$$
\end{lemma}

\section{Proofs from Section~\ref{sec::coordmeta}}
Throughout this section
we denote the players with $0, 1, \ldots, n-1$ so
that the neighbors of $i$ are $i\pm 1\bmod n$.

\subsection{\texorpdfstring{Hitting $\p$ starting from $R$ when $\Delta > \delta$}{Hitting p starting from R when Delta > delta}}
\label{app:pfromR}
In this section we prove Lemma~\ref{lem:pfromR} that provides an upper bound
on the hitting time of state $\p$ when starting from a state $\x \in R$.

\nlemma{pfromR}{\ref{lem:pfromR}}{\pfromRStm}
Remind that states in $R$ have at least two neighbors both playing $+1$
and without loss of generality we assume that $x_0=x_1=+1$.
Intuitively, for $\beta=\omega(\log n)$,
each of player $0$ and $1$ changes her strategy with very low probability.
Moreover, player $2$, when selected for update,
updates her strategy to $+1$ with high probability.
Similarly, after player $2$ has played $+1$, we have that each of
player $0, 1$ and $2$ changes her strategy with very low probability and player $3$, when selected for update,
plays $+1$ with high probability. This process repeats until every player is playing $+1$. In the following, we estimate the number of steps sufficient to have all players playing strategy $+1$ with high probability.

For sake of compactness, we will denote
the strategy of player $i$ at time step $t$ by $X_t^i$.
We start with a simple observation that lower bounds
the probability that a player picks strategy $+1$
when selected for update, given that at least one of their neighbors
is playing $+1$.
\begin{obs}
\label{obs:R:update_prob}
If player $i$ is selected for update at time $t$ then,
for $b\in\{-1,+1\}$
$$
\Prob{}{X_t^i = +1 \mid X_{t-1}^{i+b} = +1} \geq
	\left(1 - \frac{1}{1 + e^{(\Delta - \delta)\beta}}\right).
 $$
\end{obs}
We start by lower bounding the probability that
players $2, \ldots, n-1$ are selected for update at least once in this
order before  a given number $t$ of steps.
Set $\rho_1 = 0$ and, for $i=2,\dots,n-1$, let $\rho_i$ be the first time player $i$
is selected for update after time step $\rho_{i-1}$.
Thus, at time $\rho_i$ player $i$ is selected for update and players $2,\dots, i - 1$
have already been selected at least once in this order.
In particular, $\rho_{n-1}$ is the first time step at which every player $i$, $i \geq 2$,
has been selected at least once after his left neighbor.
Obviously, $\rho_i > \rho_{i-1}$ for $i=2,\ldots,n-1$.
The next lemma lower bounds the probability that $\rho_{n-1} \leq t$.
\begin{lemma}
\label{lem:R:rho_bound}
For every $\x \in R$ and every $t>0$, we have
$$
 \Prob{\x}{\rho_{n-1} \leq t} \geq 1 - \frac{n^2}{t}.
$$
\end{lemma}
\begin{proof}
Every player $i$ has probability $\frac{1}{n}$ of being selected at
any given time step.
Therefore, $\Expec{}{\rho_2} = \Expec{}{\rho_2 - \rho_1} = n$ and
$\Expec{}{\rho_i - \rho_{i - 1}} = n$,
for $i=3, \dots, n-1$. Thus, by linearity of expectation,
$$
 \Expec{}{\rho_{n-1}} = \sum_{i=2}^{n-1} \Expec{}{\rho_i - \rho_{i - 1}} \leq n^2.
$$
The lemma follows from the Markov inequality.
\end{proof}
The next lemma gives a lower bound to the probability that
$X_t^i=+1$ for all players $i$ and for $t\geq \rho_{n-1}$.
\begin{lemma}
\label{lem:R:prob0player_i}
For every $\x\in R$,
for every player $i$ and for every $t > 0$, we have
 $$
  \Prob{\x}{X_t^i = +1 \mid \rho_{n - 1} \leq t} \geq \left(1 - \frac{1}{1 + e^{(\Delta - \delta)\beta}}\right)^{t}.
 $$
\end{lemma}
% \begin{proof}
We prove the lemma first for $i\geq 2$ and then we will deal with players $0$ and $1$.
Fix player $i\geq 2$, time step $t$ and set $s_{i+1}=t$.
Starting from time step $t$ and going backward to time step $0$,
we identify the sequence of time steps $s_i>s_{i-1}>\ldots > s_2 >0$ such that,
for $j=i,i-1,\ldots,2$,
$s_j$ is the last time player $j$ has been selected before time $s_{j+1}$.
We remark that, since $t \geq \rho_{n-1} > \rho_i$ we have that players $2, \dots, i$ are selected at least once in this order and thus all the $s_j$ are well defined.
Strictly speaking, the sequence $s_i,\ldots,s_2$ depends on $i$ and $t$
and thus a more precise, and more cumbersome, notation would have been
$s_{i,j,t}$. Since player $i$ and time step $t$ will be
clear from the context, we drop $i$ and $t$.

In order to lower bound the probability that $X_t^i = +1$ for
$i \geq 2$, we first bound it in terms of the
probability that player $2$ plays $+1$ at time $s_2$
and then we evaluate this last quantity.
The next lemma is the first step of our proof.
\begin{lemma}
\label{lem:ringdom_xit}
For every $\x \in R$, every player $2\leq i \leq n-1$ and every $t \geq \rho_{n-1}$,
we have
$$
\Prob{\x}{X_{t}^i = +1 \mid \rho_{n-1} \leq t} \geq
	\left(1 - \frac{1}{1 + e^{(\Delta - \delta)\beta}}\right)^{i-2}
		\Prob{\x}{X_{s_2}^2 = +1 \mid \rho_{n-1} \leq t}.
 $$
\end{lemma}
\begin{proof}
For every $i$,
$s_i$ is the last time the player $i$ is selected for update before $t$
and thus $X_t^i = X_{s_i}^i$.
Hence, for $i=2$ the lemma obviously holds.
For $i>2$ and $j=3, \ldots, i$, we observe that, since $t \geq\rho_i$,
$s_{j-1}$ is the last time that player $j-1$ has been selected for update before time $s_{j}$
and thus $X_{s_j - 1}^{j-1} = X_{s_{j-1}}^{j-1}$. Then,
from Observation~\ref{obs:R:update_prob}, we have
 \begin{align*}
  \Prob{\x}{X_{s_j}^j = +1 \mid \rho_{n - 1} \leq t} & \geq \Prob{\x}{X_{s_j}^j = +1 \mid X_{s_j-1}^{j-1} = +1, \rho_{n - 1} \leq t}
  \cdot \Prob{\x}{X_{s_j-1}^{j-1} = +1 \mid \rho_{n - 1} \leq t}\\
  & \geq \left(1 - \frac{1}{1 + e^{(\Delta - \delta)\beta}}\right) \Prob{\x}{X_{s_{j-1}}^{j-1} = +1 \mid \rho_{n - 1} \leq t}.
 \end{align*}
We obtain the lemma by iteratively applying the same argument to $\Prob{\x}{X_{s_{j-1}}^{j-1} = +1 \mid \rho_{n - 1} \leq t}$, for $j = i-1, i-2, \ldots, 3$.
\end{proof}

We now bound the probability that player $2$ plays $+1$ at time step $s_2$.
If player $1$ has not been selected for update before time $s_2$,
then $X_{s_2-1}^1 = X_0^1 = +1$, and,
from Observation~\ref{obs:R:update_prob}, we have
\begin{equation*}
 \begin{split}
  \Prob{\x}{X_{s_2}^2 = +1 \mid \rho_{n - 1} \leq t} & \geq \Prob{\x}{X_{s_2}^2 = +1 \mid X_{s_2-1}^1 = +1, \rho_{n - 1} \leq t}\\
  & \geq \left(1 - \frac{1}{1 + e^{(\Delta - \delta)\beta}}\right).
 \end{split}
\end{equation*}
It remains to consider the case when player $1$ has been selected for update
at least once before time $s_2$.
For any fixed time step $t$,
we define a new sequence of time steps $r_0>r_1,\ldots>0$ in the following
way. We set $r_0=s_2$ and
let $r_j$, for $j>0$, be the last time player $j\mod 2$
has been selected before time $r_{j-1}$.
For the last element in the sequence, $r_k$, it holds that
player $(k+1)\mod 2$ is not selected before time step $r_k$.

Since at time $s_2$ player $2$ has been selected for update and since $r_1$ is the last time step player $1$ has been selected for update before $r_0 = s_2$, we have $X_{s_2-1}^1 = X_{r_1}^1$ and, by Observation~\ref{obs:R:update_prob},
\begin{equation}
 \label{eq:ringdom_s2}
 \begin{split}
  \Prob{\x}{X_{s_2}^2 = +1 \mid \rho_{n - 1} \leq t} & \geq \Prob{\x}{X_{s_2}^2 = +1 \mid X_{s_2-1}^1 = +1, \rho_{n - 1} \leq t}
  \cdot \Prob{\x}{X_{s_2-1}^1 = +1 \mid \rho_{n - 1} \leq t}\\
  & \geq \left(1 - \frac{1}{1 + e^{(\Delta - \delta)\beta}}\right) \cdot \Prob{\x}{X_{r_1}^1 = +1 \mid \rho_{n - 1} \leq t}.
 \end{split}
\end{equation}
Finally, we bound $\Prob{\x}{X_{r_1}^1 = +1 \mid \rho_{n - 1} \leq t}$.
\begin{lemma}
 \label{lem:R:ell_ijt}
For every $\x \in R$, time step $t$, and player $i$,
let $r_0, \ldots, r_k$ be defined as above.
If $k>0$, we have
 $$
  \Prob{\x}{X_{r_1}^{1} = +1 \mid \rho_{n-1} \leq t} \geq \left(1 - \frac{1}{1 + e^{(\Delta - \delta)\beta}}\right)^{k}.
 $$
\end{lemma}
\begin{proof}
For sake of compactness,
in this proof we denote the parity of integer $a$ with $\Par{a} = a \mod 2$.
Thus, the definition of sequence $r_j$ gives that
player $\Par{j}$ has been selected for update at time $r_j$ and
\begin{align*}
  &\Prob{\x}{X_{r_j}^{\Par{j}} = +1 \mid \rho_{n - 1} \leq t}\\
  & \qquad \qquad \qquad \qquad \geq \Prob{\x}{X_{r_j}^{\Par{j}} = +1 \mid X_{r_j-1}^{\Par{j+1}} = +1, \rho_{n - 1} \leq t}
  \Prob{\x}{X_{r_j-1}^{\Par{j+1}} = +1 \mid \rho_{n - 1} \leq t}.
 \end{align*}
If $j \neq k$ player $\Par{j+1}$ has not been selected for update between time $r_{j + 1} \leq r_j-1$ and time $r_j$ and, by Observation~\ref{obs:R:update_prob}, we obtain
\begin{equation*}
 \begin{split}
  &\Prob{\x}{X_{r_j}^{\Par{j}} = +1 \mid \rho_{n - 1} \leq t}\\
  & \qquad \qquad \qquad \qquad \geq \Prob{\x}{X_{r_j}^{\Par{j}} = +1 \mid X_{r_j-1}^{\Par{j+1}} = +1, \rho_{n - 1} \leq t}
  \Prob{\x}{X_{r_j-1}^{\Par{j+1}} = +1 \mid \rho_{n - 1} \leq t}\\
  & \qquad \qquad \qquad \qquad  \geq \left(1 - \frac{1}{1 + e^{(\Delta - \delta)\beta}}\right) \Prob{\x}{X_{r_{j+1}}^{\Par{j+1}} = +1 \mid \rho_{n - 1} \leq t}.
  \end{split}
 \end{equation*}
If $j = k$, instead, player $\Par{k+1}$ has not been selected for update before time $r_k$ and thus $X_{r_k - 1}^{\Par{k+1}} = X_0^{\Par{k+1}} = +1$.
By Observation~\ref{obs:R:update_prob}, we have
\begin{align*}
  & \Prob{\x}{X_{r_k}^{\Par{k}} = +1 \mid \rho_{n - 1} \leq t}\\
  & \qquad \qquad \qquad \qquad \geq \Prob{\x}{X_{r_k}^{\Par{k}} = +1 \mid X_{r_k-1}^{\Par{k+1}} = +1, \rho_{n - 1} \leq t} \Prob{\x}{X_{r_k-1}^{\Par{k+1}} = +1 \mid \rho_{n - 1} \leq t}\\
  & \qquad \qquad \qquad \qquad \geq \left(1 - \frac{1}{1 + e^{(\Delta - \delta)\beta}}\right). \qedhere
  \end{align*}
\end{proof}

\begin{proof}[Proof (of Lemma~\ref{lem:R:prob0player_i})]
% We are now ready to conclude the proof of Lemma~\ref{lem:R:prob0player_i}.
For every player $i \geq 2$ and $t>0$, we have
\begin{align*}
 \Prob{\x}{X_t^i = +1 \mid \rho_{n - 1} \leq t} & \geq \left(1 - \frac{1}{1 + e^{(\Delta - \delta)\beta}}\right)^{i - 2} \Prob{\x}{X_{s_2}^2 = +1 \mid \rho_{n-1} \leq t} & (\text{from Lemma~\ref{lem:ringdom_xit}})\\
 & \geq \left(1 - \frac{1}{1 + e^{(\Delta - \delta)\beta}}\right)^{i - 1} \Prob{\x}{X_{r_1}^1 = +1 \mid \rho_{n-1} \leq t} & (\text{from Equation~\ref{eq:ringdom_s2}})\\
 & \geq \left(1 - \frac{1}{1 + e^{(\Delta - \delta)\beta}}\right)^{i - 1 + k} & (\text{from Lemma~\ref{lem:R:ell_ijt}})\\
 & \geq \left(1 - \frac{1}{1 + e^{(\Delta - \delta)\beta}}\right)^t,
\end{align*}
where $k$ is the index of the last term in the sequence $r_0, r_1, \ldots$ previously defined and
where the last inequality follows from $i - 1 + k \leq t$, since the sequence of updates we are considering cannot be longer than $t$.
% This ends the proof of Lemma~\ref{lem:R:prob0player_i} for player $i\geq 2$.

The lemma for players $0$ and $1$ can be proved in a similar way.
Clearly, if player $i=0,1$ has never been selected for update before time $t$,
we have that $X_t^i = +1$ with probability $1$.
If player $i$ has been selected at least once we have to distinguish the cases $i=0$ and $i=1$. If $i=1$, we define $r_0 = t+1$ and we identify a sequence of time step $r_1 > r_2 > \ldots > 0$ as above: we have that $X_t^1 = X_{r_1}^1$ and from Lemma~\ref{lem:R:ell_ijt} follows that
$$
 \Prob{\x}{X_t^i = +1 \mid \rho_{n - 1} \leq t} \geq \left(1 - \frac{1}{1 + e^{(\Delta - \delta)\beta}}\right)^k \geq \left(1 - \frac{1}{1 + e^{(\Delta - \delta)\beta}}\right)^t,
$$
where $k$ is the last index of the sequence $r_1, r_2, \dots$.
Finally, the probability that player $0$ plays the strategy $+1$ at time $t$, given that she was selected for update at least once, can be handled similarly to the probability that player $2$ plays the strategy $+1$ at time $s_2$.
% This concludes the proof of Lemma~\ref{lem:R:prob0player_i}.
\end{proof}

The following lemma gives the probability that the hitting time of the profile $\p$ is less or equal to $t$, given that $\rho_{n-1} \leq t$.
\begin{lemma}
\label{lem:R:prob0allPlayers}
For every $\x \in R$ and every $t > 0$, we have
$$
 \Prob{\x}{\tau_{\p} \leq t \mid \rho_{n - 1} \leq t} \geq \left(1 - \frac{1}{1 + e^{(\Delta - \delta)\beta}}\right)^{nt}.
$$
\end{lemma}
\begin{proof}
%To prove our lemma we will show a bound on the probability that,
%conditioned on $\rho_{n-1} \leq t$,
%all players are playing $+1$ at time $t$.
Let $f$ be the permutation that sorts the players according to the order of last selection
for update; i.e., $f(0)$ is the last player that is selected for update,
$f(1)$ is the next to last one, and so on.
We have
\begin{align*}
 \Prob{\x}{\tau_{\p} \leq t \mid \rho_{n - 1} \leq t} & \geq \Prob{\x}{\bigwedge_{j=0}^{n-1} X_t^{f(j)} = +1 \mid \rho_{n - 1} \leq t}\\
 &= \prod_{j=0}^{n - 1} \Prob{\x}{X_t^{f(j)} = +1 \mid \bigwedge_{i=j+1}^{n-1} X_t^{f(i)} = +1, \rho_{n - 1} \leq t}\\
 &\geq \prod_{j=0}^{n - 1} \Prob{\x}{X_t^{f(j)} = +1 \mid \rho_{n - 1} \leq t}\\
 &\geq \left(1 - \frac{1}{1 + e^{(\Delta - \delta)\beta}}\right)^{nt},
\end{align*}
where for the second inequality we used that the probability of selecting $+1$ can only increase when there are other players playing this strategy, whereas the last inequality follows from Lemma~\ref{lem:R:prob0player_i}.
\end{proof}
Now we are ready to prove Lemma~\ref{lem:pfromR}.
\begin{proof}[Proof (of Lemma~\ref{lem:pfromR})]
From Lemma~\ref{lem:R:rho_bound} and Lemma~\ref{lem:R:prob0allPlayers}, we have that for every $\x \in R$ and every $t > 0$
\begin{align*}
\Prob{\x}{\tau_\p \leq t} & \geq \Prob{\x}{\rho_{n-1} \leq t}\cdot\Prob{\x}{\tau_{\p} \leq t \mid \rho_{n - 1} \leq t}\\
 & \geq \left(1 - \frac{n^2}{t}\right) \left(1 - \frac{1}{1 + e^{(\Delta - \delta)\beta}}\right)^{nt}.
\end{align*}
For sufficiently large $n$ we have that, since $\beta=\omega(\log n)$,
$$e^{(\Delta - \delta)\beta} \geq
    \frac{8-\varepsilon}{\varepsilon} \cdot
    \frac{n^3}{\log \frac{8}{8-\varepsilon}}.
$$
By setting $t=\frac{8-\varepsilon}{\varepsilon}\cdot n^2$, we have
\begin{align*}
\Prob{\x}{\tau_\p \leq t} & \geq \left(1 - \frac{\varepsilon}{8 - \varepsilon}\right)
    \left(1 - \frac{1}{1 + \frac{8-\varepsilon}{\varepsilon}\cdot\frac{n^3}{\log \frac{8}{8-\varepsilon}}}\right)^{\frac{8-\varepsilon}{\varepsilon}\cdot n^3}\\
 & \geq \frac{8\left(1 - \frac{\varepsilon}{4}\right)}{8 - \varepsilon} \frac{8 - \varepsilon}{8} = 1 - \frac{\varepsilon}{4},
\end{align*}
where the second inequality follows from the well known approximation
$1 - a \geq e^{-\frac{a}{1 - a}}$.
\end{proof}

\subsection{\texorpdfstring{Hitting $\p$ starting from $S_d$ when $\Delta > \delta$}{Hitting p starting from Sd when Delta > delta}}
\label{app:pfromSd}

In this section we prove Lemma~\ref{lem:pfromSd} that provides an upper bound
on the hitting time of state $\p$ when starting from a state $\x \in S_d$.

\nlemma{pfromSd}{\ref{lem:pfromSd}}{\pfromSdStm}

In fact, we show that from any $\x \in S_d$ the dynamics hit
after a polynomial number of steps a profile in $R$ with high probability.
Then, the aimed result easily follows by Lemma~\ref{lem:pfromR}.

Specifically, let us denote by $\negl{n}$ a function in $n$ that is smaller than the inverse of every polynomial in $n$.
Then the following upper bound on the hitting time of $R$ from the set $S_d$ holds.
\newcommand{\hitRStm}{For every $d>0$ and $\x \in S_d$, if $\Delta > \delta$, $\beta = \omega\left({\log n}\right)$ and $n$ is sufficiently large
$$\Prob{\x}{\tau_R \leq n^2} \geq \frac{2^d}{2^d+1} \left(1 - \negl{n}\right).$$}
\begin{lemma}
\label{lem:pmt:hitR}
\hitRStm
\end{lemma}

Before proving this lemma, we show how Lemma~\ref{lem:pfromSd} easily follows from Lemma~\ref{lem:pmt:hitR} and Lemma~\ref{lem:pfromR}.
\begin{proof}[Proof (of Lemma~\ref{lem:pfromSd})]
By Lemma~\ref{lem:pfromR} and Lemma~\ref{lem:pmt:hitR}, we have
\begin{align*}
 \Prob{\x}{\tau_{\p} \leq \frac{8n^2}{\varepsilon}} & \geq \Prob{\x}{\tau_{\p} \leq \frac{8n^2}{\varepsilon} \mid \tau_R \leq n^2} \Prob{\x}{\tau_R \leq n^2}\\
 & \geq \Prob{X_{\tau_R}}{\tau_{\p} \leq \frac{(8 - \varepsilon)n^2}{\varepsilon}} \Prob{\x}{\tau_R \leq n^2}\\
 & \geq \left(1 - \frac{\varepsilon}{4}\right) \frac{2^d}{2^d + 1} \left(1 - \negl{n}\right) \geq \frac{2^d}{2^d+1} - \frac{\varepsilon}{4}. \qedhere
\end{align*}
\end{proof}

We now prove Lemma~\ref{lem:pmt:hitR}.
Specifically we will bound $\Prob{\x}{\tau_{R} \leq t \wedge \tau_R \leq \tau_\m}$ and will use the fact that
\begin{equation}
 \label{eq:hitR_and_m}
 \Prob{\x}{\tau_R \leq t} \geq \Prob{\x}{\tau_{R} \leq t \wedge \tau_R \leq \tau_\m}.
\end{equation}
Let $\theta^\star$ be the first time at which all players have been selected at least once.
We define players playing $+1$ in profile $\x$ as the \emph{plus-players} of $\x$ and their neighbors as \emph{border-players}.
Consider the event $E$ that ``a border-player is selected for update before at least one of her neighboring plus-players''
and denote by $\overline{E}$ its complement.
Observe that
\begin{equation}
 \label{eq:hitR_eq}
 \begin{aligned}
  \Prob{\x}{\tau_R \leq t \wedge \tau_R \leq \tau_\m} & \geq \Prob{\x}{\tau_R \leq t \wedge \tau_R \leq \tau_\m \mid E \wedge \theta^\star \leq t} \Prob{\x}{E \wedge \theta^\star \leq t}\\
  & = \Prob{\x}{\tau_R \leq t \wedge \tau_R \leq \tau_\m \mid E \wedge \theta^\star \leq t} \Prob{\x}{E \mid \theta^\star \leq t} \Prob{\x}{\theta^\star \leq t}.
 \end{aligned}
\end{equation}

We next bound the three components appearing in the last line of \eqref{eq:hitR_eq}.
A bound on the probability that $\theta^\star \leq t$ directly follows from the coupon collector argument;
we include a proof of this bound for completeness.
\begin{lemma}
 \label{lem:theta_star}
 For every $t > 0$,
 $$
  \Prob{\x}{\theta^\star \leq t} \geq 1 - n e^{-t/n}.
 $$
\end{lemma}
\begin{proof}
The logit dynamics at each time step select a player for update uniformly and independently of the previous selections. Thus, the probability that $i$ players are never selected for update in $t$ steps is $\left(1 - \frac{i}{n}\right)^t$ and
\[
 \Prob{\x}{\theta^\star > t} \leq \sum_{i = 1}^{n - 1} \left(1 - \frac{i}{n}\right)^t \leq \sum_{i = 1}^{n - 1} e^{-\frac{it}{n}} \leq n e^{-t/n}. \qedhere
\]
\end{proof}

Next we bound the probability of the event $E$ given that all players have been selected at least once.
\begin{lemma}
\label{lemma:eq:tau_finite:R}
For every $\x\in S_d$, $d > 0$ and every $t > 0$
$$
 \Prob{\x}{E \mid \theta^\star \leq t} \geq \frac{2^d}{2^d+1}.
$$
\end{lemma}
\begin{proof}
Let us define $\ell(\x)$ as the number of border-players in $\x$. Note that, since there are no adjacent plus-players in $\x$, $\ell(\x) \geq d$.

The proof proceeds by induction on $d$.
Let $d=1$ and denote by $i$ the {plus}-player.
Since we are conditioning on $\theta^\star \leq t$,
all players are selected at least once by time $t$
and thus the probability that one of
the two neighbors of $i$ is selected for update before $i$ is selected
is $\frac{2}{3} = \frac{2^d}{2^d+1}$.

Suppose now that the claim holds for $d-1$ and consider $\x \in S_d$.
Denote by $T_\x$ the set of all the {plus}-players in $\x$
and their {border}-players and let $i$ be the first player in $T_\x$
to be selected for update
(notice that $i$ is well defined since $\theta^\star \leq t$).
Observe that, if $i$ is a {border}-player, then the event $E$ occur and
this happens with probability $\frac{l(\x)}{l(\x)+d}$.
If $i$ is a plus-player, we consider the subset
$\overline{T}_\x \subset T_\x$ of the remaining $d-1$ {plus}-players and
their {border}-players.
The event $E$ will occur if and only if at least one {border}-player in
$\overline{T}_\x$ is selected before one of its neighboring {plus}-players.
Notice though that $\overline{T}_\x = T_\y$,
for $\y \in S_{d-1}\setminus R$ such that $y_i = -1$ and
$\y_{-i} = \x_{-i}$. Thus, by inductive hypothesis,
$\Prob{\y}{E \mid \theta^\star \leq t} \geq \frac{2^{d-1}}{2^{d-1}+1} = \frac{2^{d}}{2^{d}+2}$.
Thus,
\begin{align*}
 \Prob{\x}{E \mid \theta^\star \leq t} & = \frac{l(\x)}{l(\x)+d}+\frac{d}{l(\x)+d} \cdot \Prob{\y}{E \mid \theta^\star \leq t}\\
 & \geq \frac{l(\x)}{l(\x)+d}+\frac{d}{l(\x)+d} \cdot \frac{2^d}{2^d+2}\\
 & = 1 - \frac{2d}{(l(\x)+d)(2^d + 2)}\\
 & \geq 1 - \frac{1}{2^d + 2},
\end{align*}
where the last inequality follows from $l(\x)\geq d$.
The claim follows since $1 - \frac{1}{2^d + 2} \geq \frac{2^d}{2^d+1}$.
\end{proof}
Finally suppose that the event $E$ occurs, i.e., there is a border-player $i$ selected for update before than at least one of the neighboring plus-players, and $\theta^\star \leq t$. Let $\tau$ be the time step at which $E$ occurs. Note that, since all players have been selected at least once before $t$, then $\tau < t$. Moreover, at time $\tau$, the player $i$ has at least one neighbor playing strategy $+1$, and thus she will play this strategy with probability at least $\left(1 - \frac{1}{1 + e^{(\Delta - \delta)\beta}}\right)$. Then we have
\begin{equation}
\label{eq:boundR_cond}
 \Prob{\x}{\tau_R \leq t \wedge \tau_R \leq \tau_\m \mid E \wedge \theta^\star \leq t} \geq \Prob{\x}{X_{\tau}^i = +1 \mid E \wedge \theta^\star \leq t} \geq \left(1 - \frac{1}{1 + e^{(\Delta - \delta)\beta}}\right),
\end{equation}
where $X_{\tau}^i$ denotes the strategy of $i$ in the profile $X_\tau$.

Now we are ready to prove Lemma~\ref{lem:pmt:hitR}.
\begin{proof}[Proof (of Lemma~\ref{lem:pmt:hitR})]
From \eqref{eq:hitR_and_m}, \eqref{eq:hitR_eq}, \eqref{eq:boundR_cond}, Lemma~\ref{lem:theta_star} and Lemma~\ref{lemma:eq:tau_finite:R} we have
\[
 \Prob{\x}{\tau_R \leq n^2} \geq \left(1 - \frac{1}{1 + e^{(\Delta - \delta)\beta}}\right) \cdot \frac{2d-1}{2d} \cdot \left(1 - n e^{-n}\right) = \frac{2d-1}{2d} \cdot (1 - \negl{n}),
\]
where the last equality holds for $n$ sufficiently large since $\beta = \omega(\log n)$.
\end{proof}

\subsection{\texorpdfstring{Hitting $\{\p, \m\}$ starting from $S_d$ when $\Delta > \delta$}{Hitting \{p, m\} starting from Sd when Delta > delta}}
\label{apx:pmfromSd}
In this section we prove Lemma~\ref{lem:pmfromSd} that provides an upper bound
on the hitting time  of the set $\{\p,\m\}$ when starting from a state $\x \in S_d$.

\nlemma{pmfromSd}{\ref{lem:pmfromSd}}{\pmfromSdStm}

The proof is similar to the proof of Lemma \ref{lem:pfromSd}, except that now
we need a bound on the probability to hit $R$ \emph{or} $\m$
in polynomial time when starting from a state in $S_d$.
Next lemma gives us such a bound.
\begin{lemma}
\label{lem:pmt:hitRm}
For all $\x \in \{\pm 1\}^n$,
if $\Delta > \delta$, $\beta = \omega\left({\log n}\right)$
and $n$ is sufficiently large
$$
 \Prob{\x}{\tau_{R \cup \{\m\}} \leq n^2} \geq 1 - \negl{n}.
$$
\end{lemma}

Lemma~\ref{lem:pmfromSd} then immediately follows.
\begin{proof}[Proof (of Lemma~\ref{lem:pmfromSd})]
By Lemma~\ref{lem:pfromR} and Lemma~\ref{lem:pmt:hitRm}, we have
\begin{align*}
 \Prob{\x}{\tau_{\{\p,\m\}} \leq \frac{8n^2}{\varepsilon}} & \geq \Prob{\x}{\tau_{\{\p,\m\}} \leq \frac{8n^2}{\varepsilon} \mid \tau_{R \cup \{\m\}} \leq n^2} \Prob{\x}{\tau_{R \cup \{\m\}} \leq n^2}\\
 & \geq \Prob{X_{\tau_{R \cup \{\m\}}}}{\tau_{\{\p,\m\}} \leq \frac{(8 - \varepsilon)n^2}{\varepsilon}} \Prob{\x}{\tau_{R \cup \{\m\}} \leq n^2}\\
 & \geq \left(1 - \frac{\varepsilon}{4}\right) \left(1 - \negl{n}\right) \geq 1 - \frac{\varepsilon}{2}. \qedhere
\end{align*}
\end{proof}

Let us now prove Lemma~\ref{lem:pmt:hitRm}. First of all, we note that
\begin{equation}
\label{eq:hitRm}
 \Prob{\x}{\tau_{R \cup \{\m\}} \leq t} = \Prob{\x}{\tau_{R} \leq t \wedge \tau_R \leq \tau_\m} + \Prob{\x}{\tau_{\m} \leq t \wedge \tau_\m \leq \tau_R}.
\end{equation}
A lower bound on the first term on the right hand side of the above equation was given in previous section. In fact,
from \eqref{eq:hitR_eq}, \eqref{eq:boundR_cond} and Lemma~\ref{lem:theta_star}, we have
\begin{equation}
\label{eq:hitRforRm}
 \Prob{\x}{\tau_{R} \leq t \wedge \tau_R \leq \tau_\m} \geq \left(1 - \frac{1}{1 + e^{(\Delta - \delta)\beta}}\right) \cdot \left(1 - n e^{-t/n}\right) \cdot \Prob{\x}{E \mid \theta^\star \leq t}.
\end{equation}
Similarly,
\begin{equation}
\label{eq:hitmforRm}
\begin{aligned}
 \Prob{\x}{\tau_\m \leq t \wedge \tau_\m \leq \tau_R} & \geq \Prob{\x}{\tau_\m \leq t \wedge \tau_\m \leq \tau_R \mid \overline{E} \wedge \theta^\star \leq t} \Prob{\x}{\overline{E} \wedge \theta^\star \leq t}\\
 & = \Prob{\x}{\tau_\m \leq t \wedge \tau_\m \leq \tau_R \mid \overline{E} \wedge \theta^\star \leq t} \Prob{\x}{\overline{E} \mid \theta^\star \leq t} \Prob{\x}{\theta^\star \leq t}\\
 & \geq \left(1 - n e^{-t/n}\right) \cdot \Prob{\x}{\tau_\m \leq t \wedge \tau_\m \leq \tau_R \mid \overline{E} \wedge \theta^\star \leq t} \Prob{\x}{\overline{E} \mid \theta^\star \leq t}
\end{aligned}
\end{equation}
where the last inequality follows from Lemma~\ref{lem:theta_star}.

Next lemma bounds $\Prob{\x}{\tau_\m \leq t \wedge \tau_\m \leq \tau_R \mid \overline{E} \wedge \theta^\star \leq t}$.
\begin{lemma}
\label{lem:probTauM}
For every $d \geq 0$, for every $\x \in S_d$ and every $t > 0$, we have
$$
 \Prob{\x}{\tau_\m \leq t \wedge \tau_\m \leq \tau_R \mid \overline{E} \wedge \theta^\star \leq t} \geq 1 - \frac{t}{e^{2\delta\beta}}.
$$
\end{lemma}
\begin{proof}
Suppose that the event $E$ does not occur, i.e., all plus-players are selected before their neighboring border-players, and $\theta^\star \leq t$. Let $\tau$ be the time step at which the last plus-player is selected. Note that, since all players have been selected at least once before $t$, then $\tau < t$. Thus, if in the first $\tau$ time steps the selected player adopts strategy $-1$, then $\tau_{\m} \leq \tau \leq t$ and $\tau_\m \leq \tau_R$, i.e., by denoting the player selected for update at time step $j$ as $i(j)$,
\begin{align*}
 \Prob{\x}{\tau_{\m} \leq t \wedge \tau_\m \leq \tau_R \mid \overline{E} \wedge \theta^\star \leq t} & \geq \Prob{\x}{\bigwedge_{j=1}^\tau X_j^{i(j)} = -1 \mid \overline{E} \wedge \theta^\star \leq t}\\
 & = \prod_{j=1}^{\tau} \Prob{\x}{X_j^{i(j)} = -1 \mid \overline{E} \wedge \theta^\star \leq t \wedge \bigwedge_{k=1}^j X_k^{i(k)} = -1}.
\end{align*}
Given that no border-player is selected before the corresponding plus-player and that at each previous time step the selected player has adopted strategy $-1$, we have that, for every $j$, player $i(j)$ has both neighbors playing $-1$. Hence, $i(j)$ adopts strategy $-1$ with probability $\left(1 - \frac{1}{1 + e^{2\delta\beta}}\right)$.
Thus,
$$
  \Prob{\x}{\tau_{\m} \leq t \wedge \tau_\m \leq \tau_R \mid \overline{E} \wedge \theta^\star \leq t} \geq \left(1 - \frac{1}{1 + e^{2\delta\beta}}\right)^\tau \geq \left(1 - \frac{1}{1 + e^{2\delta\beta}}\right)^t \geq \left(1 - \frac{t}{e^{2\delta\beta}}\right),
$$
where the last inequality follows from the approximations $1 - a \leq e^{-a}$ and $1 - a \geq e^{-\frac{a}{1 - a}}$ for $0 \leq a \leq 1$.
\end{proof}

Now we are ready to prove Lemma~\ref{lem:pmt:hitRm}.
\begin{proof}[Proof (of Lemma~\ref{lem:pmt:hitRm})]
From \eqref{eq:hitRforRm}, \eqref{eq:hitmforRm} and Lemma~\ref{lem:probTauM}, for $n$ sufficiently large and $\beta = \omega(\log n)$ we have both
$$
 \Prob{\x}{\tau_{R} \leq n^2 \wedge \tau_R \leq \tau_\m} \geq \left(1 - \negl{n}\right) \cdot \Prob{\x}{E \mid \theta^\star \leq n^2}.
$$
and
$$
 \Prob{\x}{\tau_\m \leq n^2 \wedge \tau_\m \leq \tau_R} \geq \left(1 - \negl{n}\right) \cdot \Prob{\x}{\overline{E} \mid \theta^\star \leq n^2}.
$$
Then, from \eqref{eq:hitRm} it follows that
\[
 \Prob{\x}{\tau_{R \cup \{\m\}} \leq n^2} \geq \left(1 - \negl{n}\right) \left(\Prob{\x}{E \mid \theta^\star \leq n^2} + \Prob{\x}{\overline{E} \mid \theta^\star \leq n^2}\right) = 1 - \negl{n}. \qedhere
\]
\end{proof}

\subsection{\texorpdfstring{Hitting $\{\p, \m\}$ when $\Delta = \delta$}{Hitting \{p, m\} when Delta = delta}}
\label{app:proofsw}
In this section we prove Lemma~\ref{lem:ringnodom_taupm} that provides an upper bound
on the hitting time  of the set $\{\p,\m\}$ when $\Delta = \delta$.

\nlemma{ringnodom_taupm}{\ref{lem:ringnodom_taupm}}{\rndtaupmStm}

Let us start by introducing some useful notation.
We say that the profile $\x$ has a \emph{plus-block} of size $l$ starting at player $i$ if
$x_i=x_{i+1}=\ldots=x_{i+l-1}=+1$ and $x_{i-1}=x_{i+l}=-1$ and players $i$ and $i+l-1$ are the \emph{border} players of the block.
A similar definition is given for \emph{minus-blocks}.
Notice that every profile $\x\ne\p,\m$ has the same number of {plus}-blocks and {minus}-blocks and
this number is called the \emph{level} of $\x$ and is denoted by $\ell(\x)$. We set $\ell(\p)=\ell(\m)=0$.
%
%The following observation gives the \emph{level structure} of the potential function (note that we are studying the case $\Delta=\delta$).
%\begin{obs}
% \label{obs:ringnodom:pot_level}
%For every profile $\x$, the potential of $\x$ is $\Pot(\x) =(n-2\ell(\x))\Delta$, regardless of the sizes of the plus-blocks and minus-blocks.
%\end{obs}
Moreover, for a profile $\x$, we defines $s_+(\x)$ as the number of {plus}-blocks of size 1, $s_-(\x)$ as the number of {minus}-blocks of size 1 and set $s(\x)=s_+(\x) + s_-(\x)$.

% \subsection{Hitting $\{\p, \m\}$ when starting from $\x$}
We would like to study how long it takes to the logit dynamics to hit $\p$ or $\m$. Let $\x$ be the starting profile of the logit dynamics and assume that $\ell(\x) = \ell$.
Since $\p$ and $\m$ have level $0$, the dynamics have to go down $\ell$ levels before hitting one of these two target profiles.
To reach a profile in a smaller level the dynamics have to reach
a profile having a monochromatic block of size 1 is reached, select the unique player of this block for update and have this player changing
her strategy to the same strategy as her neighbors.
Let us denote with $\tau_i$ the hitting time of a profile at level $i$, i.e., $\tau_i = \tau_{\{\x \colon \ell(\x) = i\}}$.
By linearity of expectation, we have that
\begin{equation}
 \label{eq:hit_bound_sum_levels}
 \Expec{\x}{\tau_{\{\p, \m\}}} \leq \sum_{i=0}^{\ell-1} \max_{\x : \ell(\x) = i+1} \Expec{\x}{\tau_i}.
\end{equation}
Thus, to bound the hitting time of $\{\p, \m\}$ we have only to compute $\Expec{\x}{\tau_i}$ for any level $i = 0, 1, \cdots, \ell-1$ and for any profile $\x$ such that $\ell(\x) = i+1$.

Fix $\x$ be a profile of level $i+1$ and number arbitrarily its $2(i+1)$ monochromatic blocks. We denote by $k_j(\x)$ the size of the
$j$-th monochromatic block.
For each $1 \leq j \leq 2(i+1)$, let us define the quantity $\theta_{j}$ as follows.
Suppose that the dynamics starting from $\x$ hit for the first time a profile of level $i$ after $t$ steps and this happens because the $j$-th monochromatic block disappears; then, we set $\theta_{j}=t$ and $\theta_{j'}=+\infty$ for all $j'\ne j$.
Given that the starting profile $\x$ is at level $i+1$, we have $\tau_i = \min_j \theta_{j}$ and, thus,
$$ \Expec{\x}{\tau_i}=\Expec{\x}{\min_j \theta_{j}}\leq
	\max_j \Expec{\x}{\theta_{j} \mid \theta_{j} < \theta_{j'} \text{ for all } j'\neq j}.$$
In order to have a more compact notation we define
$$
 \gamma_{i,l} = \max_{1 \leq j \leq 2(i + 1)} \max_{\substack{\x \colon \ell(\x)=i+1\\k_j(\x)=l}} \Expec{\x}{\theta_{j} \mid \theta_{j} < \theta_{j'} \text{for all } j'\neq j},
$$
and set $\gamma_i=\max_l\gamma_{i,l}$. Observe that
\begin{equation}
 \label{eq:tau_i_gamma_i}
 \Expec{\x}{\tau_i}\leq\gamma_i
\end{equation}
and it is not hard to see that $\gamma_{i,l}$
is non-decreasing with $l$. Next lemma gives a bound $\gamma_i$ in terms of $\gamma_{i+1}$. Using recursively this lemma we will be able to upper bound $\gamma_i$ for any $i \geq 0$.
\begin{lemma}
 \label{lem:ringnodom_gammabound}
For $0 \leq i < \lfloor n/2 \rfloor - 1$
$$
 \gamma_i \leq n^3 \left(1 + \frac{1}{1+e^{2\Delta\beta}} \gamma_{i + 1}\right).
$$
Moreover, $\gamma_{\lfloor n/2 \rfloor - 1} \leq n^3$.
%  where $b_i = n + \frac{n - 4(i+1) + s(\x^\star)}{1+e^{2\Delta\beta}} \gamma_{i+1}$.
\end{lemma}
\begin{proof}

We start by bounding $\gamma_{i,l}$ for any $l$. Let $\x$ and $j$ be the profile and the monochromatic block that attain the maximum $\gamma_{i,l}$,
respectively.  We remark that we are conditioning on the event that $\theta_{i,j}<\theta_{i,j'}$  for all $j'\neq j$ (i.e., the $j$-th block is
the first one to disappear). To bound $\gamma_{i,l}$ we distinguish three cases, depending on the value of $l$.

\smallskip \noindent \underline{$l=1$:}
Let $u$ be the unique player of the $j$-th monochromatic block and let $v$ be the player selected for update. Consider all the possible selections of $v$:
\begin{itemize}
\item
If $u=v$ and $v$ changes her strategy the block disappears and thus $\theta_j = 1$. This event occurs with probability $\frac{1}{n}\cdot\left(1 -\frac{1}{1+e^{2\Delta\beta}}\right)$.

\item Suppose that $v$ is a neighbor of $u$ and she changes her strategy.
Observe that $v$ cannot belong to a monochromatic block of size 1 since we are assuming that $j$ is the first block to disappear.
Thus, the two neighbors of $v$ are playing different strategies and when $v$ has to change her strategy she selects at random.
We can conclude that the probability that $v$ is selected and she changes her strategy is at most $1/2\cdot 2/n=1/n$. Moreover, after this update the dynamics reach a profile of level $i+1$ where the size of the $j$-th block increases to $2$.

\item Suppose that $v$ is not a border player of any monochromatic block and she changes her strategy.
The players that are not at the border of a block are $n-4(i+1)+s(\x)$ and each of them changes her strategy with probability $\frac{1}{1+e^{2\Delta\beta}}$. We can conclude that the probability of such an event is $\frac{n-4(i+1)+s(\x)}{n(1+e^{2\Delta\beta}})$ and in this case the dynamics reach a profile of level $i + 2$.

\item for all the remaining choices for $v$ both the level of the reached profile and the length of the $j$-th monochromatic block remain the same as in the starting profile.
\end{itemize}
Summing over all the possible choices for $v$ and observing that $\gamma_{i,2}\geq\gamma_{i,1}$ we have
\begin{align*}
\gamma_{i,1} & \leq \frac{1}{n} \left(1 - \frac{1}{1+e^{2\Delta\beta}}\right) + \frac{1}{n} (1 + \gamma_{i,2}) + \frac{n-4(i+1) + s(\x)}{n} \frac{1}{1+e^{2\Delta\beta}} (1 + \gamma_{i+1})\\
 & + \left(\frac{n - 2}{n} - \frac{n - 4(i +1) + s(\x) - 1}{n} \frac{1}{1 + e^{2\Delta\beta}}\right) (1 + \gamma_{i,1}).
\end{align*}
By simple calculations and using the fact that $n - 4(i+1) + s(\x) \geq 0$, we obtain
$$
 \gamma_{i,1} \leq \left(\frac{1}{2} + \frac{1}{4e^{2\Delta\beta} + 2}\right) \left(n + \gamma_{i,2} + \frac{n - 4(i + 1) + s(\x)}{1+e^{2\Delta\beta}} \gamma_{i + 1}\right).
 $$
Finally, since $\left(\frac{1}{2} + \frac{1}{4e^{2\Delta\beta} + 2}\right) \leq \frac{2}{3}$ for $\beta=\omega(\log n)$, we can conclude that
\begin{equation}
 \label{eq:ringnodom_gamma1}
 \gamma_{i,1} \leq \frac{2}{3} (\gamma_{i,2} + b_i),
\end{equation}
where $b_i = n + \frac{\gamma_{i+1}}{1+e^{2\Delta\beta}} \cdot \max_{\y \colon \ell(\y) = i +1} (n - 4(i+1) + s(\y))$.

\smallskip \noindent \underline{$1<l<n-2i-1$:}
Let $v$ be the player selected for update. Consider all the possible selections of $v$:
\begin{itemize}
 \item suppose $v$ is one of the two players at the border of the $j$-th monochromatic block and she changes her strategy.
 Since the border players are two and their neighbors are playing different strategies, $v$ selects her new strategy at random and thus the probability of this event is $1/n$. Moreover, the dynamics reach a profile $\y$ of level $i+1$ where the length of the $j$-th monochromatic block decreases to $l-1$.

\item suppose $v$ does not belong to the $j$-th monochromatic block but she is a neighbor of its border players
and she changes her strategy. Then the dynamics reach a profile $\y$ of level $i+1$.
Since $v$ cannot belong to a monochromatic block of size 1 (otherwise her block would disappear before block $j$)
we can state that the two neighbors of $v$ are playing different strategies. Thus, when $v$ updates her strategy
she chooses each of the two alternatives with probability $1/2$.
Since there are two players adjacent to the border players of block $j$,
this case happens with probability at most $1/n$.

\item suppose $v$ is a player that is not a broader player of any monochromatic block and she changes her strategy.
Then, when $v$ changes her strategy a monochromatic block is split and the reached profile has level $i+2$.
Notice that there are $n - 4(i+1) + s(\x)$ such players and each of them has probability
$\frac{1}{1+e^{2\Delta\beta}}$ to change her strategy.

\item for all the remaining choices for $v$ both the level of the reached profile and the length of the $j$-th monochromatic block remain the same as in the starting profile.
\end{itemize}
Hence,
\begin{align*}
 \gamma_{i,l} & \leq \frac{1}{n} (1 + \gamma_{i,l-1}) + \frac{1}{n} (1 + \gamma_{i,l+1}) + \frac{n - 4(i + 1) + s(\x)}{n} \frac{1}{1+e^{2\Delta\beta}} (1 + \gamma_{i+1})\\
 & + \left(\frac{n - 2}{n} - \frac{n - 4(i +1) + s(\x)}{n} \frac{1}{1 + e^{2\Delta\beta}}\right) (1 + \gamma_{i,l}).
\end{align*}
By simple calculations, similar to the ones for the case $l=1$, we obtain
$$
 \gamma_{i,l} \leq \frac{1}{2} (\gamma_{i,l-1} + \gamma_{i,l+1} + b_i).
$$
From the previous inequality and Equation~\ref{eq:ringnodom_gamma1},
a simple induction on $l$ shows that, for every $1\leq l< n-2i-1$, we have
\begin{equation}
 \label{eq:ringnodom_backrecur}
 \gamma_{i,l} \leq \frac{1}{l + 2} \left((l + 1)\gamma_{i,l+1} + \frac{l(l + 3)}{2} b_i\right).
\end{equation}
Moreover, from Equation~\ref{eq:ringnodom_backrecur},
we can use a simple inductive argument to show that, for every $h \geq 1$,
\begin{equation}
 \label{eq:ringnodom_forrecur}
\begin{aligned}
\gamma_{i,l} & \leq \frac{l + 1}{l + h + 1} \gamma_{i, l + h} + \frac{l+1}{2} b_i \sum_{j=l}^{l + h - 1} \frac{j(j+3)}{(j+1)(j+2)} \\
 & \leq \frac{l + 1}{l + h + 1} \gamma_{i,l + h} + \frac{l+1}{2} hb_i.
\end{aligned}
\end{equation}

\smallskip \noindent \underline{$l=n-2i-1$: }
in this case all blocks other than the $j$-th have size 1. Thus,
every time one of these players is selected for update she doesn't change her strategy
otherwise there would be a monochromatic block disappearing before block $j$.
This means that the size of the $j$-th monochromatic block cannot increase.
By using an argument similar to the one used in the previous cases, we obtain that
$$
 \gamma_{i,n - 2i - 1} \leq \gamma_{i,n - 2i - 2} + b_i.
$$
By using Equation~\ref{eq:ringnodom_backrecur}, we have
$$
 \gamma_{i,n - 2i - 1} \leq \frac{(n - 2i - 2)(n - 2i + 1) + 2(n - 2i)}{2} b_i \leq \frac{n^2}{2} b_i.
$$
Finally, for every $l\geq 1$, by using Equation~\ref{eq:ringnodom_forrecur} with $h = n - 2i - 1 - l$, we have
\[
 \gamma_{i,l} \leq \frac{l + 1}{n - 2i} \gamma_{i,n - 2i - 1} + \frac{(l + 1)(n - 2i - 1 - l)}{2} b_i \leq n^2 b_i.
\]
The lemma finally follows by observing that $\max_{\y \colon \ell(\y) = i +1} (n - 4(i+1) + s(\y)) \leq n$ for $0 \leq i < \lfloor n/2 \rfloor - 1$ and it is exactly 0 for $i = \lfloor n/2 \rfloor - 1$.
\end{proof}
\begin{cor}\label{cor:ringnodom_gammabound}
If $\beta=\omega(\log n)$, then for every $i \geq 0$,
$\gamma_i = \OO(n^3).$
\end{cor}
\begin{proof}
From Lemma~\ref{lem:ringnodom_gammabound} we have
$\gamma_{\lfloor n/2 \rfloor - 1} \leq n^3$.
Instead, for $0 \leq i < \lfloor n/2 \rfloor - 1$, we have
\begin{equation}
 \label{eq:ringnodom_gammafinal}
 \gamma_i \leq n^3 \left(1 + \frac{1}{1+e^{2\Delta\beta}} \gamma_{i + 1}\right) \leq n^3 \left(1 + \sum_{j=1}^{\lfloor n/2 \rfloor-i-1} \left(\frac{n^3}{1+e^{2\Delta\beta}}\right)^j\right).
\end{equation}
The corollary follows by observing that, if $\beta = \omega\left(\frac{\log n}{\Delta}\right)$, then the summation in Equation~\ref{eq:ringnodom_gammafinal} is $o(1)$.
\end{proof}

The above corollary gives
a polynomial bound to the time that the dynamics take to go from a profile at level $i+1$  to a profile at level $i$.
Lemma~\ref{lem:ringnodom_taupm} easily follows.

\begin{proof}[Proof (of Lemma \ref{lem:ringnodom_taupm})]
From \eqref{eq:hit_bound_sum_levels} and \eqref{eq:tau_i_gamma_i}, for every profile $\x$ at level $1\leq \ell \leq n/2$ we have
$$
 \Expec{\x}{\tau_{\p,\m}} \leq \sum_{i=0}^{\ell-1} \gamma_i = \OO(n^4),
$$
where the last bound follows from Corollary \ref{cor:ringnodom_gammabound}.
The lemma then follows from the Markov inequality.
\end{proof}

\subsection{\texorpdfstring{Bounding $\alpha_\x$ when $\Delta = \delta$}{Bounding alpha-x when Delta = delta}}
\label{apx:alpha}
Here we prove the following lemma.
\begin{lemma}
 \label{lem:ringnodom_conv}
For every $d\geq 0$, every profile $\x$ with exactly $d$ players playing $+1$ and $\beta = \omega(\log n)$
$$\alpha_\x = \frac{d}{n} \pm o(1).$$
\end{lemma}
\begin{proof}
Trivially, $\alpha_\p = 1$ and $\alpha_\m = 0$.
We next show that for  $\beta = \omega(\log n)$  and $\x$ with exactly $d$ players playing $+1$,
$\alpha_\x = \frac{d}{n} + \lambda_\x$, for some $\lambda_\x = o(1)$.
By the definition of Markov chains we know that
$$\alpha_\x=P(\x,\x)\cdot \alpha_\x+\sum_{y\in N(\x)} P(\x,\y)\cdot p_\y.$$
We then partition the neighborhood $N(\x)$ of profile $\x$ of level $i$
in $5$ subsets, $N_1(\x)$, $N_2(\x)$, $N_3(\x)$, $N_4(\x)$, $N_5(\x)$
such that, for two profiles $\y_1,\y_2$ in the same subsets it holds that $P(\x,\y_1)=P(\x,\y_2)$.
Then
\begin{itemize}
\item
$N_1(\x)$ is the set of profiles $\y$ obtained from $\x$ by changing the
strategy of a player of a plus-block of size 1.
Observe that $|N_1(\x)|=s_+(\x)$. Moreover, for every $\y \in N_1(\x)$, $\y$ is at level $i-1$,
has $d - 1$ players playing $+1$ and $P(\x,\y) = \frac{1}{n}\cdot (1-\frac{1}{1 + e^{2\Delta\beta}})$.

\item
$N_2(\x)$ is the set of profiles $\y$ obtained from $\x$ by changing the
strategy of a player of a minus-block of size 1.
Observe that $|N_2(\x)|=s_-(\x)$. Moreover, for every $\y \in N_2(\x)$, $\y$ is at level $i-1$,
has $d + 1$ players playing $+1$ and $P(\x,\y) = \frac{1}{n}\cdot (1-\frac{1}{1 + e^{2\Delta\beta}})$.

\item
$N_3(\x)$ is the set of profiles $\y$ obtained from $\x$ by changing the
strategy of a border player of a plus-block of size greater than 1.
Observe that  $|N_3(\x)|=2(i - s_+(\x))$. Moreover, for every $\y \in N_3(\x)$, $\y$ is at level $i$,
and has $d - 1$ players playing $+1$ and $P(\x,\y)=1/2n$.

\item
$N_4(\x)$ is the set of profiles $\y$ obtained from $\x$ by changing the
strategy of a border player of a minus-block of size greater than 1.
Observe that  $|N_4(\x)|=2(i - s_-(\x))$.
Moreover, for every $\y \in N_4(\x)$, $\y$ is at level $i$,
and has $d + 1$ players playing $+1$ and $P(\x,\y)=1/{2n}$.

 \item
$N_5(\x)$ is the set of all the profiles $\y\in N(\x)$ that do not belong to any of the previous $4$ subsets.
Observe that  $|N_5(\x)|=n-4i+s(\x)$.
Moreover, for every $\y \in N_5(\x)$, $\y$ is at level $i+1$,
and $P(\x,\y)=\frac{1}{n}\cdot\frac{1}{1+e^{2\Delta\beta}}$.
\end{itemize}
Moreover, we have that
$$P(\x, \x) = \frac{s(\x)}{n} \frac{1}{1 + e^{2\Delta\beta}} + \frac{2i - s(\x)}{n} + \frac{n - 4i + s(\x)}{n} \left(1 - \frac{1}{1 + e^{2\Delta\beta}}\right).$$
 Then, we have
\begin{equation}
\label{eq:ringnodom_px0_eq}
\begin{aligned}
 \alpha_\x
 & = \frac{1}{n}\left(1 - \frac{1}{1 + e^{2\Delta\beta}}\right) \left(\sum_{\y \in N_1(\x)} \alpha_\y + \sum_{\y \in N_2(\x)} \alpha_\y\right) + \frac{1}{2n} \left(\sum_{\y \in N_3(\x)} \alpha_\y + \sum_{\y \in N_4(\x)} \alpha_\y\right) \\
 & \quad + \frac{1}{n}\frac{1}{1 + e^{2\Delta\beta}} \sum_{\y \in N_5(\x)} \alpha_\y\\
 & \quad + \left(\frac{s(\x)}{n} \frac{1}{1 + e^{2\Delta\beta}} + \frac{2i - s(\x)}{n} + \frac{n - 4i + s(\x)}{n} \left(1 - \frac{1}{1 + e^{2\Delta\beta}}\right)\right) \alpha_\x\\
 & = \frac{1}{n} \left(\sum_{\y \in N_1(\x)} \alpha_\y + \sum_{\y \in N_2(\x)} \alpha_\y\right) + \frac{1}{2n} \left(\sum_{\y \in N_3(\x)} \alpha_\y + \sum_{\y \in N_4(\x)} \alpha_\y\right) + \frac{n - 2i}{n} \cdot \alpha_\x + \frac{c}{1 + e^{2\Delta\beta}},
\end{aligned}
\end{equation}
where
$$
 c=\frac{1}{n}\left(\sum_{\y \in N_5(\x)} \alpha_\y - \sum_{\y \in N_1(\x) \cup N_2(\x)} \alpha_\y - (n - 4i) \alpha_\x\right).
$$
We notice that, since $1 \leq i \leq n/2$ and $|N_1(\x)|+|N_2(\x)|,|N_5(\x)| \leq n$, we have $|c|\leq 2$ and thus the last term in Equation~\ref{eq:ringnodom_px0_eq} is negligible in $n$ (since $\beta=\omega(\log n)$).
Hence we have that the following condition holds for every level $i\geq 1$ and every profile $\x$ at level $i$:
$$
 \alpha_\x = \frac{1}{2i} \left(\sum_{\y \in N_1(\x)} \alpha_\y + \sum_{\y \in N_2(\x)} \alpha_\y\right) + \frac{1}{4i} \left(\sum_{\y \in N_3(\x)} \alpha_\y + \sum_{\y \in N_4(\x)} \alpha_\y\right) + \eta_\x,
$$
where $\eta_\x$ is negligible in $n$.
This gives us a linear system of equations in which the number of equations is the same that the number of variables.

Let us consider the \emph{polished} version of this system, in which we omit the negligible part in every equation. We will compute a solution of the polished system. Finally, we argue the solution of original system should be close to the polished one.

Let us denote with $\alpha^\star_\x$ the variable depending on $\x$ in the polished system. Note that for each level $i$ and every profile $\x$ at level $i$, $\alpha^\star_\x$ does not depend on profiles at higher level. This enable us to compute a solution for the polished system inductively on level $i$.
Indeed, for every profile $\x$ at level $0$
(this is only possible for $d=0$ or $d=n$),
 we have, as discussed above, $\alpha^\star_\x = \frac{d}{n}$.
Now, consider $\x$ at level $i$ with exactly $d$ players playing $+1$.
By assuming that, for every profile $\y$ at level $i-1$,
$\alpha^\star_\y = \frac{d_\y}{n}$, where $d_\y$ is the number of player playing $+1$ in $\y$,
we have
\begin{equation}
\label{eq:ringnodom_cond}
 \alpha^\star_\x = \frac{s_+(\x)}{2i} \cdot \frac{d -1}{n} + \frac{s_-(\x)}{2i} \cdot \frac{d +1}{n} +
\frac{1}{4i} \left(\sum_{\y \in N_3(\x)} \alpha^\star_\y + \sum_{\y \in N_4(\x)} \alpha^\star_\y\right).
\end{equation}
Let us now consider Equation~\ref{eq:ringnodom_cond} for each $\x$ at level $i$.
These gives another linear system of equations.
This system has a unique solution:
indeed, the number of equations and the number of variables coincides and, moreover, the matrix of coefficients is a diagonally dominant matrix (since $|N_3(\x)| + |N_4(\x)| \leq 4i$) and thus it is non-singular.
We will show this solution must set $\alpha^\star_\x = \frac{d}{n}$ for every profile $\x$ at level $i$ with exactly $d$ players playing $+1$.
Indeed, with this assignment the right hand side of
the Equation~\ref{eq:ringnodom_cond} becomes
$$
 \frac{s_+(\x)}{2i} \frac{d -1}{n} + \frac{s_-(\x)}{2i} \frac{d +1}{n} + \frac{i - s_+(\x)}{2i} \frac{d -1}{n} + \frac{i - s_-(\x)}{2i} \frac{d +1}{n}
 = \frac{d}{n}.
$$
Thus, we can conclude $\alpha_\x^\star = \frac{d}{n}$ for each profile $\x$ with exactly $d$ players playing $+1$.

Now, let $\lambda_\x = \left| \alpha_\x - \alpha_\x^\star \right|$.
As $n$ grows unbounded, any original equation approaches the polished one,
so we need $\alpha_\x$ to approach $\alpha_\x^\star$.
Then $\lambda_\x = o(1)$ for every profile $\x$.
\end{proof}

\end{document}